\documentclass[10pt,pre,twocolumn,nofootinbib]{revtex4-1}

\pdfoutput=1 
\usepackage{aas_macros}
\usepackage{hyperref}
\usepackage{amsmath}
\usepackage{grffile}
\usepackage{graphicx}   
\usepackage{verbatim}   
\usepackage{color}      
\usepackage{subfigure}  
\usepackage{hyperref}   
\newcommand{\borg}{{\tt BORG}}
\newcommand{\tmpp}{{2M++}}

\begin{document}

\title{The primordial magnetic field in our cosmic backyard}
\author{Sebastian Hutschenreuter}
\altaffiliation{Max Planck Institute for Astrophysics, Karl-Schwarzschildstr.1, 85741 Garching, Germany}
\altaffiliation{Ludwig-Maximilians-Universit\"at M\"unchen, Geschwister-Scholl-Platz 1, 80539 Munich, Germany}
\author{Sebastian Dorn}
\altaffiliation{Max Planck Institute for Astrophysics, Karl-Schwarzschildstr.1, 85741 Garching, Germany}
\altaffiliation{Ludwig-Maximilians-Universit\"at M\"unchen, Geschwister-Scholl-Platz 1, 80539 Munich, Germany}
\author{Jens Jasche}
\altaffiliation{Department of Physics, Stockholm University, Albanova University Center, SE 106 91
Stockholm, Sweden}
\author{Franco Vazza}
\altaffiliation{University of Bologna, Department of Physics and Astronomy, Via Gobetti 93/2, I-40129, Bologna, Italy}
\altaffiliation{Universit\"{a}t Hamburg, Hamburger Sternwarte, Gojenbergsweg 112, 40129, Hamburg, Germany}
\altaffiliation{Istituto di Radioastronomia, INAF, Via Gobetti 101, 40129 Bologna, Italy}
\author{Daniela Paoletti}
\altaffiliation{INAF/OAS Bologna, Via Gobetti 101, I-40129 Bologna, Italy}
\altaffiliation{INFN, Sezione di Bologna, via Irnerio 46, I-40127 Bologna, Italy}
\author{Guilhem Lavaux}
\altaffiliation{Sorbonne Universit\'e \& CNRS, UMR 7095, Institut
d'Astrophysique de Paris, 98 bis bd Arago, 75014 Paris, France}
\author{Torsten A. En{\ss}lin}
\altaffiliation{Max Planck Institute for Astrophysics, Karl-Schwarzschildstr.1, 85741 Garching, Germany}
\altaffiliation{Ludwig-Maximilians-Universit\"at M\"unchen, Geschwister-Scholl-Platz 1, 80539 Munich, Germany}

\begin{abstract}
We reconstruct for the first time the three dimensional structure of magnetic fields on cosmological scales, which were seeded by density perturbations during the radiation dominated epoch of the Universe and later on were evolved by structure formation. To achieve this goal, we rely on three dimensional initial density fields inferred from the 2M++ galaxy compilation via the Bayesian {\borg} algorithm. Using those, we estimate the magnetogenesis by the so called Harrison mechanism.
This effect produced magnetic fields exploiting the different photon drag on electrons and ions in vortical motions, which are exited due to second order perturbation effects in the Early Universe. 
Subsequently we study the evolution of these seed fields through the non-linear cosmic structure formation by virtue of a MHD simulation to obtain a 3D estimate for the structure of this primordial magnetic field component today. 
At recombination we obtain large scale magnetic field strengths around  $10^{-23} \mathrm{G}$, with a power spectrum peaking at about $ 2\, \mathrm{Mpc}^{-1}h$ in comoving scales. At present we expect this evolved primordial field to have strengths above $\approx 10^{-27}\, \mathrm{G}$ and $\approx 10^{-29}\, \mathrm{G}$ in clusters of galaxies and voids, respectively. We also calculate the corresponding Faraday rotation measure map and show the magnetic field morphology and strength for specific objects of the Local Universe. These results provide a reliable lower limit on the primordial component of the magnetic fields in these structures. 
\end{abstract}

\maketitle

\section{Introduction}
\label{intro}
Inference of primordial magnetic fields opens a unique window into the Early Universe between inflation and recombination. Although a variety of different astrophysical processes may generate magnetic fields, the primordial magnetic seed may very well be the origin of observed magnetic fields in galaxies and clusters. Primordial magnetic fields are a viable candidate for the $10^{-16}\,\mathrm{G}$ to $\approx\, 10^{-15}\,\mathrm{G}$ \cite{neronov, neronov2, tavecchio,dolag} fields expected in cosmic voids due to the non-observation of GeV emission from TeV blazars among other explanations \citep{pfrommer}. In any case, they represent by definition the minimal amount of magnetic fields present in the Universe.
Literature provides a variety of very diverse effects for the generation of primordial magnetic fields coherent on a large range of scales. A incomplete list of possible magnetogenesis effects may include mechanisms at the end of inflation (e.g. during the reheating phase or exploiting the electroweak phase transitions), during QCD phase transitions or effects that make use of speculative non-standard model physics such as gravitational coupling of the gauge potential or string theory effects. Very often these mechanisms struggle with producing the necessary field strengths and/or, especially the post inflationary models, the necessary coherence lengths for large-scale magnetic fields. The scale problem might be solved, at least for helical magnetic fields, via an inverse cascade which transfers magnetic power to larger scales \citep{savliev12, savliev13}. Recent works have shown that a similar mechanism exists for non-helical fields \citep{Brandenburg_nonhelical}, although the process is still poorly understood \citep{reppin}. For a further discussion on different magnetogenesis models, we refer the reader to comprehensive review articles \citep{kandus,subramanian,yamazaki,durrer}.     \par  
A more conservative Ansatz solely relying on the assumption of a $\Lambda$CDM Universe and conventional plasma physics was proposed by Matarrese et al. \citep{mat}. This approach is based on a mechanism initially proposed by Harrison \citep{harrison} in 1970. During the later phases of the radiation dominated epoch of the Universe a two fluid battery effect occurred between the proton fluid and the tightly coupled electron-photon fluid. The densities $\rho_{(\alpha)}$, with $\alpha \in \left\lbrace m,\gamma \right\rbrace$ for baryons and the electron-photon fluid respectively, of the two components scale with the scale factor $a(t)$ as $\rho_{(\mathrm{m})} \sim a(t)^{-3}$  and $\rho_{(\gamma)} \sim a(t)^{-4}$, respectively. Therefore,  the separately conserved angular momenta $L_{(\alpha)} \sim \rho_{(\alpha)}\,\omega_{(\alpha)}\, r^5$ in a rotational setup with radius $r(t) \sim a(t)$  requires the angular velocities $\omega_{(\alpha)}$ to depend on $a(t)$ with $\omega_{(\mathrm{m})} \sim a^{-2}$ and $\omega_{(\gamma)} \sim a^{-1}$, respectively. In other words, protons spin down faster than electrons, as the latter are carried by the still dominant photons. This difference in rotation then leads to currents that induce magnetic fields \citep{harrison}. 
The necessary vortical motion of both proton and radiation fluid are caused by effects that can be expressed as second order perturbations of the fluid equation \cite{mat}. \par
The recent progress on the inference of the actual 3D realization of the large-scale dark matter structure and its formation history in the Local Universe and the fact that the Harrison mechanism is solely founded on well established plasma physics allows us to calculate the seed magnetic fields that had to be generated by this effect as well as their present day morphology and strength. Since these fields have to exist today in combination with fields of other sources, we are therefore able to provide credible lower bounds on the primordial magnetic field strength in the Nearby Universe. We structure this article as following:\par
Section \ref{sec:theory} first describes the outcome of the Matarrese paper \citep{mat} and then presents the computational steps that take us from dark matter over densities to magnetic fields. 
Section \ref{sec:data} gives a short overview on the dark matter density reconstruction used in this work.
Section \ref{sec:results} provides the intermediate results on magnetic field configuration and power spectrum at radiation matter equality.
Section \ref{sec:mhd} shows the results of the subsequent MHD simulation.
Section \ref{sec:outlook} contains a summary and an outlook on potential improvements.
 
\section{Theory}
\label{sec:theory}
This paper strongly relies on the theoretical framework outlined by Matarrese et al. \citep{mat}. This approach describes primordial density perturbations in the Early Universe before the recombination epoch as sources of magnetic fields via the Harrison mechanism \citep{harrison}.\par
In the first part of this section we will summarize their assumptions and results. The second part describes the implemented reconstruction approach to translate our knowledge on dark matter over-densities into magnetic field estimates, first described by Dorn \citep{dorn}.  
\subsection{Basics}

\begin{figure}[t]
\begin{tabular*}{0.75\columnwidth}{|@{\extracolsep{\fill} }c|c c|}
\hline
Parameter & Value & \\
\hline
$H_0$ & 67.74 $\mathrm{km}\,\mathrm{Mpc}^{-1}\mathrm{s}^{-1}$ & \\ 
$h$ & 0.6774 & \\
$\Omega_{\Lambda}$ & $0.6911$ & \\
$\Omega_{m}$ & $0.3089$ &\\
$\Omega_{r}$ & $5.389 \cdot 10^{-5}$ &\\
$\Omega_{k}$ & $0.0$ &\\
$z_{\mathrm{eq}}$ & $3371$ &\\
\hline
\end{tabular*}
\caption{\label{tab:param}Table of cosmological parameters used in this work \citep{Planck15_param}.}
\end{figure}

All calculations here are performed using the standard $\Lambda$CDM model assuming the cosmological parameters described in Tab.\ref{tab:param} following the 2015 results of the Planck mission \citep{Planck15_param}. \newline
Following \citep{mat} we further assume that the dominant constituents of the Universe in the relevant time frame behave as perfect fluids of dark matter, electrons, protons and photons. All equations and calculations are performed in Poisson gauge with the following line element:
\begin{eqnarray}
ds^2 = & a^2(\eta)\,(-(1+2\phi)\,d\eta^2 + 2 \chi_i\, d\eta \,dx^i\nonumber\\ & + \left( \left( 1-2\psi\right) \delta_{ij} + \chi_{ij}\right)dx^idx^j)
\label{metric}
\end{eqnarray}
\newline
$a$ is the scale factor depending on conformal time $\eta$. $\phi$ and $\psi$ are the Bardeen potentials, $\chi_i$ and $\chi_{ij}$ are vector and tensor perturbations, respectively. \newline
The perfect fluid assumption results in a vanishing anisotropic stress tensor, which yields $\phi = \psi \equiv \varphi$ to first order in perturbation theory. \par
As perfect fluids are assumed, the energy momentum tensor simplifies to 
\begin{equation}
T^{(\alpha)\mu}_{\nu} = (P^{(\alpha)} + \rho^{(\alpha)})\,u^{(\alpha)\mu}u_{(\alpha)\nu} + P^{(\alpha)}\delta^{\mu}_{\nu}\;\;.
\end{equation}
with $P^{(\alpha)}$ the pressure, $\rho_{\alpha}$ the density and $u_{\alpha}^{\mu}$ the bulk velocity for each component $\alpha$.
Pressure and density of a component are related via an equation of state
\begin{equation}
\label{eq:state}
P^{(\alpha)} = w^{(\alpha)}\,\rho^{(\alpha)}\;.
\end{equation}
We define the energy over-density with respect to the mean energy density $\overline{\rho^{(\alpha)}}$ (which is the same quantity as $\rho^{(\alpha)}_0$ in \citep{mat}) of a component as
\begin{equation}
\delta^{(\alpha)} = \frac{\rho^{(\alpha)}}{\overline{\rho}^{(\alpha)}} - 1 \;.
\end{equation}
All quantities ($\delta^{(\alpha)},u^{(\alpha)},\varphi,\chi_i,\chi_{ij}$) can now be perturbed up to second order and related via their respective momentum equations including source terms to describe interactions. The coupling between the baryonic and radiation components is assessed by a tight coupling approximation to zeroth order which implies $ v_i^{(p)} \approx v_i^{(e)} \approx v_i^{(\gamma)}$. This sets Thomson and Coulomb interaction terms to zero in this order. The curl of the momentum equations for the proton and photon components of the fluid gives evolution equations for the respective vorticities. The magnetic fields will be generated by vortical structures in the conductive non-relativistic baryonic component. To understand this, however, we turn our eye to the dominating photon component in that fluid. If a fluid component $\alpha$ was considered separately, its vorticity $\omega^{(\alpha)}$ is a conserved quantity as stated by Kelvin's circulation theorem. This holds for each order of perturbation theory separately, in particular for the dominating photons:

\begin{equation}
\label{eq:kelvin}
\omega^{\prime(\gamma)} = 0.
\end{equation} 
Given that we expect no vorticity in the initial conditions, external sources are absent and we have an ideal fluid where pressure and density gradients are aligned, photon vorticity will always be zero. 
There is however a subtelty that that comes into play due to the fact that photons experience pressure.
The photon vorticity equation in second order is \citep{mat}

\begin{eqnarray}
\label{eq:photon_vort}
\omega^{(\gamma)}_{i\,(2)} = &  -\frac{1}{2a^2}\epsilon_{ijk}\left[ av^{j,k\,(\gamma)}_{(2)} + a\chi^{j,k}_{(2)} \right.\\ & \left. + v^{j\,(\gamma)}_{(1)}\varphi^{,k}_{(1)} + v^{\prime j\,(\gamma)}_{(1)}v^{k\,(\gamma)}_{(1)} \right] \nonumber
\end{eqnarray}

The vorticity of photons in second order perturbation is not equal to the curl of the second order perturbed velocity field, but includes coupled first order terms. Since we need to obey the conservation law in Eq. \eqref{eq:kelvin} and these first order terms are non-zero, a curl in the photon velocity field needs to be induced. If we now turn our eye to the proton vorticity equation these squared first order terms are absent due to the vanishing pressure:

\begin{equation}
\label{eq:proton_vort}
 \omega^{(\mathrm{p})}_{i\,(2)} = -\frac{1}{2a}\epsilon_{ijk}\left[ v^{j,k\,(\mathrm{p})}_{(2)} + \chi^{j,k}_{(2)}\right] 	 
\end{equation}

The crucial connection is now that the tight coupling of the fluid components does not couple the vorticities of protons and photons but their velocities. Therefore the right part in Eq. \eqref{eq:proton_vort} is non-zero and acts as a external source term for vorticity. Connecting this with Maxwells equations, we get an equation for the generation of magnetic fields. In other words the arising proton vorticity needs to be offset by an external (magnetic) force, in order to keep angular momentum conserved. The tight coupling approximation is discussed in more detail in \ref{subsec:simplifications}. \par
Matarrese et al. \cite{mat} expressed the evolution equation for the magnetic field in terms of the first order scalar perturbations of the metric $\varphi_{(1)}$ which in the Newtonian limit gives the gravitational potential. By assuming negligible resistivity and therefore omitting magnetic diffusion terms and performing (at least partially) an analytic integration, they get \citep{mat, dorn}   
\begin{eqnarray}
\label{eq:B_eq}
B = & -
\frac{m_p}{e\,a\mathcal{H}^2}\;
\left[2\nabla\varphi^\prime\times\nabla\varphi -\frac{1}{12\mathcal{H}} \nabla\left(\Delta\varphi\right)\times\nabla\varphi \right. \nonumber \\ & -\left.  \frac{1}{12\mathcal{H}^2} \nabla\left(\Delta\varphi^\prime\right)\times\nabla\varphi \right] \nonumber\\& - \frac{1}{a^2} \int_{\eta_I}^\eta d\tilde{\eta}\,\frac{a}{\mathcal{H}}\,\nabla\varphi^\prime\times \nabla \varphi + \frac{a_I^2}{a^2}B_I 
\end{eqnarray}
for the magnetic field at time $\eta$, assuming some initial field $B_I$  at time $\eta_I$. 
A prime denotes derivation by conformal time. $m_p$ is the proton mass, $e$ the elementary charge, $a$ the scale factor and $\mathcal{H} = a^\prime / a $ the comoving Hubble constant. 
In the formulation above, the generation of magnetic fields is the result of a coupling between first order temporal and spatial gradients of the scalar perturbations. In other words, the generation of magnetic fields is the result of dynamics in the gravitational potential, which in turn are a result of the gravitational pull on infalling matter through the horizon and the counteracting radiation pressure. This close connection to the Baryon Accoustic Oscillations (BAOs) will be evident in the power spectra of our results at recombination. Even for $\varphi^{\prime}=0$, which is true in the matter dominated era, this terms is not zero, as the second term only depends on spatial gradients. The formulation above is very convenient, as the sole dependence on the scalar perturbations makes the connection 
to initial conditions and the corresponding state of the Universe today very easy, as will be demonstrated in the next section. 
Furthermore all terms in Eq. \eqref{eq:B_eq} include at least two derivatives of $\varphi$. For large scales above the horizon scales this implies a scaling of the magnetic field power spectrum of approximately $k^4$. Similar arguments where brought up in \citep{fenu}. This is also in accordance with the causality limit on uncorrelated magnetic fields, which demands at least a $k^2$ scaling \citep{durcap}.  
The integral term was omitted by Mataresse in their analysis on the correlation structure of the field. The initial field $B_I$ can safely be set to zero due to the $a_I^2/a^2$ factor. The assumption of small resistivity can be justified via considering the diffusion timescale 
\begin{equation}
\tau_{\mathrm{diff}} = 4\pi\sigma L^2,
\end{equation} 
where $L$ typical scale of magnetic structures and $\sigma$ is the electron conductivity. Electron momentum transfer is dominated via Thomson scattering, we can therefore write $\sigma=\frac{n_e e^2}{n_\gamma m_e \sigma_T}$. Plugging everything in, using the cosmological parameters from the Planck mission \citep{Planck15_param} e.g. at recombination and $L \approx 1\,\mathrm{Mpc}$, yields
\begin{equation}
\tau_{\mathrm{diff}} \approx 10^{42}\, \mathrm{s}, 
\end{equation} 
which is orders of magnitude higher than the age of the Universe at $t_\mathrm{U} \approx  4.4 \cdot 10^{17} \mathrm{s}$. Therefore neglecting the diffusion term is justified. In general, this is true throughout the history of the Universe, at least after inflation and on large scales \citep{turner}. High conductivity also implies flux freezing, which will lead to field amplification during structure formation in the late time evolution of the magnetic field as we will see in our results.
\par 
In the following we discuss the implementation of equation \eqref{eq:B_eq}. 

\subsection{Implementation}

We now need to calculate $\varphi$ and its spatial and temporal derivatives with respect to conformal time. We begin our calculation by translating the CDM density perturbations measured\footnote{For further comments on the data see section \ref{sec:data} and \citep{Jasche14}} shortly before the last scattering surface $ \delta_{\mathrm{cdm}}(z \approx 1000)$  into primordial initial conditions deep inside the radiation epoch at $z_{\mathrm{p}}$. 
Magnetogenesis can only take place on scales which have entered the cosmic horizon at the corresponding epoch. Therefore it makes sense to make this the criterion for $z_{\mathrm{p}}$. We know that the horizon condition can be roughly written as
\begin{equation}
\label{eq:imp_horizon}
k_{\mathrm{h}}\cdot \eta_{\mathrm{h}} \approx 1 
\end{equation} 
with the conformal time measured in units of one over length, the speed of light set to one and with $ \eta $ indicating conformal time.
Knowing that the smallest scales of the grid correspond to $k_{256} \approx  2.39\, h\,\mathrm{Mpc}^{-1}$ and $k_{512} \approx  4.78\, h\,\mathrm{Mpc}^{-1}$ with respective grid sizes of 256 and 512 points respectively, we know that the initial times must be on the order of the grid resolution $\eta_{256} \approx 0.42\, \mathrm{Mpc}\,h^{-1}$ and $\eta_{512} \approx 0.21\,  \mathrm{Mpc}\,h^{-1}$ with the speed of light set to one. The equivalent redshifts are $z_{256} \approx 9.7\cdot 10^5$ and $z_{512} \approx 1.9\cdot 10^6$. Finally, $z_{\mathrm{p}} = 10^7$ was adopted in this work, as it safely satisfies the aforementioned condition.
We obtain $\delta_{\mathrm{cdm}}(z_{\mathrm{p}}) $ by using linear cosmological transfer functions and calculate the total energy over-densities $\delta_{\mathrm{tot}}(z_{\mathrm{p}}) $ from it,

\begin{equation}
\delta_{\mathrm{tot}}(z_{\mathrm{p}}) = \frac{4}{3}\delta_{\mathrm{cdm}}(z_{\mathrm{p}}) = \frac{4}{3} T(k,z_{\mathrm{p}},z_{\mathrm{rec}})\,\delta_{\mathrm{cdm}}(z_{\mathrm{rec}})
\label{eq:transfer_eq}
\end{equation}

The $4/3$ factor comes from the adiabatic super-horizon solutions for the density perturbations (see e.g. \cite{mabert}). The transfer functions $T(k,z_{\mathrm{p}},z_{\mathrm{rec}}) $ in Poissonian gauge were calculated by the CLASS code \cite{class}.
They contain all relevant physics at linear order. \newline
The peculiar gravitational potential $\varphi(k,\eta)$ in the radiation epoch (implying $ w^{(\gamma)} \approx 1/3 $) evolves as (see, e.g., \citep{mat})
\begin{equation}
\varphi(k,\eta) = \frac{3\,j_1(x)}{x} \varphi_0(k)
\label{eq:phi_eq}
\end{equation}  
in Fourier space with initial conditions $\varphi_0$ at redshift $z = 10^7$, $j_1$ is the spherical Bessel function of first order and 
\begin{equation}
x = \frac{k\,\eta}{\sqrt{3}}.
\end{equation} 
Furthermore, the linearised Einstein equations relate the total energy perturbations to the potential by

\begin{equation}
\delta_{\mathrm{tot}} =  \frac{2}{3\mathcal{H}^2}\left(\Delta \varphi - 3 \mathcal{H}\left[\varphi^\prime + \mathcal{H}\varphi\right] \right).
\label{eq:delta_eq}
\end{equation}

With this equation we can calculate the initial $\varphi_0$ in Fourier representation as

\begin{equation}
\varphi_0(k) = \frac{3\,\mathcal{H}^2}{2k^2\,\frac{j_1(x)}{x}-6\mathcal{H}\left[\frac{\partial}{\partial\,\eta}\left(\frac{j_1(x)}{x}\right)+\mathcal{H}\frac{j_1(x)}{x}\right]}\,\delta_{\mathrm{tot}}.
\label{eq:phi_0_eq}
\end{equation}

From there on we can use Eq. \eqref{eq:phi_eq} again to calculate the potential and its derivatives at any time up to $z_{\mathrm{eq}}$.\\
The potential $\varphi$ will tend to a constant after radiation-matter equality, as pressure becomes negligible. This means that the first and third term in Eq. \eqref{eq:B_eq} do not contribute to magnetogenesis from the epoch of radiation-matter equality to recombination. We therefore evaluate these terms at radiation matter equality ($z\approx 3371$). From there on, these magnetic field terms are then propagated to recombination at redshift $z = 1088$ via the induction equation of magneto-hydrodynamics (MHD) (assuming again perfect conductivity):

\begin{equation}
\frac{\partial B}{\partial \eta} = \nabla \times (v \times B)
\label{eq:induction}
\end{equation}

The fluid velocity $v$ is also calculated in first order perturbation theory.
The second term of equation \eqref{eq:B_eq} contains no time derivative of the potential and is therefore evaluated at recombination ($z=1088$).
\par
\begin{center}
\begin{figure}
\includegraphics[page =2,scale=0.38]{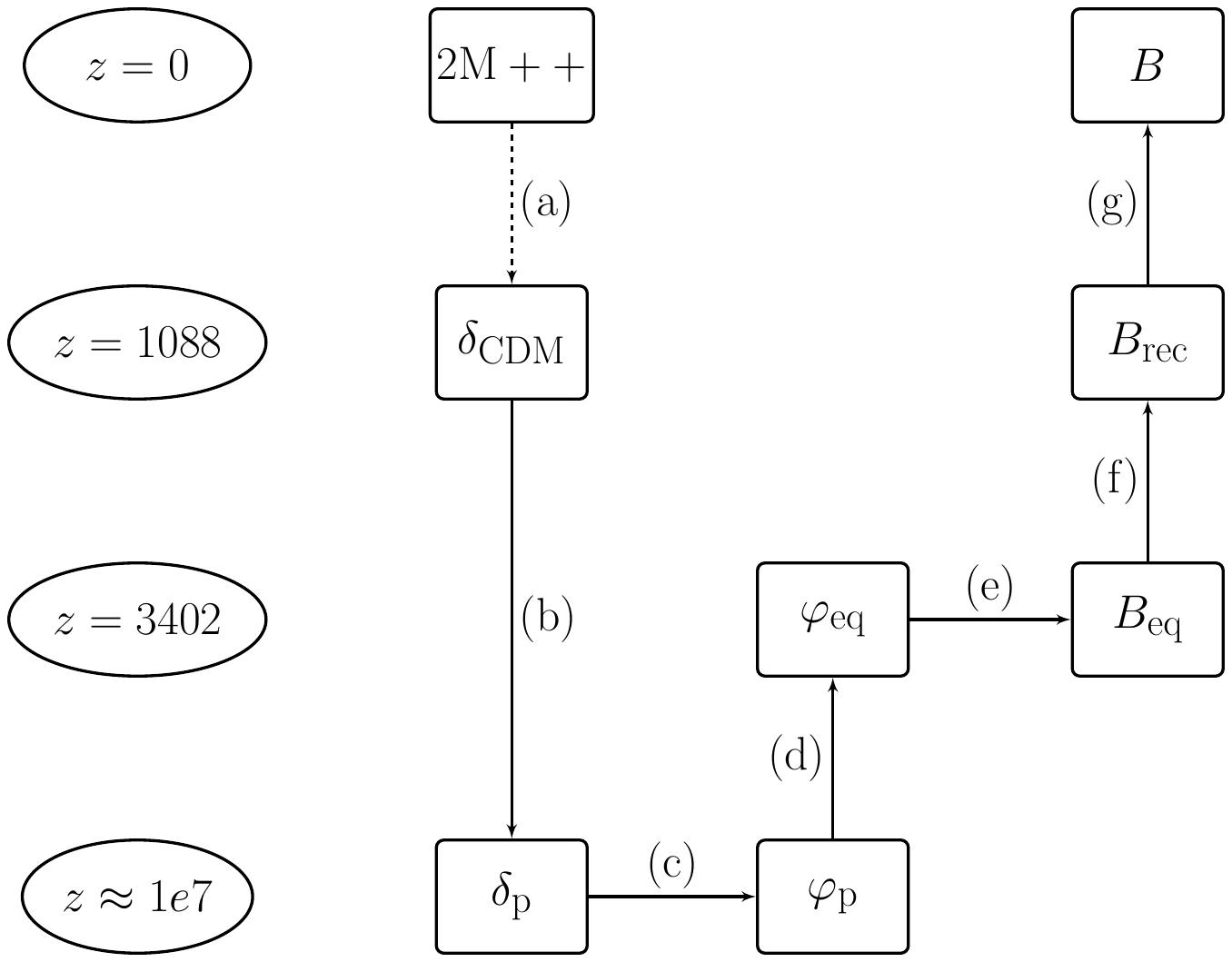}
\caption{An illustration of the implemented algorithm. The ellipses indicate relevant redshifts. The labels near the arrows refer to the following steps: (a) Dark matter inference from galaxy data via {\borg}, (b) Linear dark matter transfer functions (Eq. \ref{eq:transfer_eq}), (c) translation of dark matter density to potential $\varphi$ (Eq. \ref{eq:phi_0_eq}), (d) time evolution of the potential via Eq. \ref{eq:phi_eq}, (e) calculation of the magnetic field (Eq. \ref{eq:B_eq}), (f) induction equation (Eq. \ref{eq:induction}), (g) full MHD solver (see section \ref{sec:mhd})}
\label{fig:flow}
\end{figure}
\end{center}

 
We illustrate the steps of the calculation in Fig. \ref{fig:flow}. 

\subsection{Simplifications}
\label{subsec:simplifications}

This calculation contains simplifying assumptions to keep the evolution equation for the magnetic field anaytically solvable. For completeness, those shall be discussed here.
\par
The evolution of the potential via Eq. \eqref{eq:phi_eq} is performed for a radiation dominated Universe with equation of state \eqref{eq:state}. The transition to the matter dominated era is modelled in an abrupt way with $w = \frac{1}{3}$ before equality and $w = 0$ afterwards. As the real transition is smooth, scales in the order of the equality horizon maybe affected by the modelling and magnetogenesis may even take place even after recombination. A heuristic modelling via e.g. hyperbolic functions was not performed as the additional time dependence makes the evolution equations for the potential not analytically solvable. This could be incorporated in the model if needed for the price of more contrieved calculations. 
Related to that, the coupling between electrons and photons is modelled via the tight coupling approximation as mentioned earlier. Thomson scattering is very efficient for scales larger than the mean free path of the photons, which at recombination can be estimated via
\begin{equation}
\label{eq:thomson}
d_{\mathrm{Thomson}} = \frac{1}{n_e a \sigma_T} \approx \, 2\, \mathrm{Mpc}\,h^{-1}
\end{equation}
in comoving scales. As we will see in the next section, our calculation is performed on a $\approx 1.3 \mathrm{Mpc}\,h^{-1}$ grid. Therefore the tight coupling should ideally be expanded to higher order in case of the Thomson coupling.
\par
The above mentioned shortcomings where overcome by more detailed studies on the the generation of primordial magnetic fields via the Harrison mechanism, which have been conducted by several authors in the past 15 years. Gopal et al. \citep{Sethi} showed that differences between electron and photon velocities lead to source terms for magnetic field generation. Saga et al. \citep{Saga} and Fenu et al. \citep{fenu} have refined this calculation by including anisotropic stresses stemming from the imperfect Thomson coupling. A similar calculation was done by Christopherson et al. \citep{christ} first on the generation of vorticity in second order and later on the subsequent magnetic field generation \citep{nalson} via the introduction of non-adiabatic pressure terms. \par 
All of these approaches give source terms on which the Harrison mechanism can operate. The corresponding equations can in principle all be solved given suitable initial conditions. Only the approach shown above, however, gives an analytically integrable expression. All other models require the iterative solution of the respective magnetic field evolution equation in combination with all relevant quantities throughout the whole plasma era of the Universe up until recombination. This requires considerable computational effort to be implemented on the $512^3$ voxel grid used in this work. For this reason, and as the field strengths which were found in \citep[][]{Sethi, Saga, nalson, fenu, fidler} are comparable to the ones found by Mataresse \citep{mat}, we find the above mentioned simplifications acceptable. Fidler et al.\citep{fidler} show that the exact treatment of the Thomson coupling gives rise to significant magnetogenesis even after last scattering, which highlights that a better modelling around recombination would be desirable, given that one can shoulder the resulting computational complications.

\section{Data}
\label{sec:data}

This work builds upon three dimensional dark matter
density fields previously inferred from the {\tmpp} galaxy
compilation \citep{Lavaux2mpp} via the BORG algorithm \citep{Lavaux}. The {\borg} algorithm is a full scale Bayesian inference framework aiming
at the analysis of the linear and mildly-non-linear regime
of the cosmic large scale structure (LSS) \citep[][]{Jasche13,Jasche14}. In particular it performs dynamical LSS inference from galaxy redshift surveys employing a second order Lagrangian perturbation model. As such the {\borg} algorithm naturally
accounts for the filamentary structure of the cosmic web
typically associated to higher order statistics as induced
by non-linear gravitational structure formation processes.
A particular feature, relevant to this work, is the ability of
the {\borg} algorithm to infer Lagrangian initial conditions
from present observations of the galaxy distribution \citep[][]{Jasche13,Jasche14,Lavaux}. More specifically the algorithm explores a LSS posterior distribution consisting of a Gaussian
prior for the initial density field at a initial scale factor
of a = 0.001 linked to a Poissonian likelihood model of galaxy formation at redshift z = 0 via a second order Lagrangian
perturbation theory (2LPT) model [for details see \citep[][]{Jasche13,Jasche14,Lavaux}], that is conditioned to the {\tmpp} galaxy compilation \citep{Lavaux}. Besides typical observational systematics and uncertainties, such as survey geometries, selection functions
and noise this algorithm further accounts for luminosity
dependent galaxy bias and performs automatic noise calibration \citep{Lavaux}. The {\borg} algorithm accounts for all joint
and correlated uncertainties in inferred quantities by performing a Markov Monte Carlo chain in multi-million
dimensional parameter spaces via an efficient implementation of a Hamiltonian Monte Carlo sampler \citep{Jasche13}. As a result the algorithm provides a numerical representation of the LSS posterior in the form of data constrained realizations of the present three dimensional dark matter distribution and corresponding initial conditions from which it formed. It is important to remark that each individual Markov sample qualifies for a plausible realisation of the LSS.
Each sample of the dark matter distribution consists of a box with $256^3$ grid points and $677.7\, \mathrm{Mpc}\,h^{-1}$ edge length, resulting in a resolution of approximately $2.5\, \mathrm{Mpc}\,h^{-1}$. For one sample of \texttt{BORG} we increase the resolution of the grid to $512^3$ by augmenting the large scale modes with random fluctuations consistent with the known dark matter power spectrum. This sample is then propagated into todays configuration via a MHD simulation as explained in the following section.  
As described above we now apply the Harrison mechanism on data constrained initial conditions of the Nearby Universe.

\section{MHD simulations}
\label{sec:mhd}
The MHD computation is started from the magnetic field generated at $z=1088$ and is
evolved to $z=0$ using the cosmological code \texttt{ENZO} \citep[][]{enzo14}. 
\texttt{ENZO} is a grid based code that follows the dynamics
of dark matter with a particle-mesh N-body method, and a combination of 
several possible  shock-capturing Riemann solvers to evolve the gas
component \citep[][]{enzo14}. The MHD method employed in this paper is the 
Dedner ``cleaning" method \citep[][]{ded02}, which makes use of  hyperbolic
divergence cleaning
to keep the (spurious) divergence of the magnetic field as low as possible during the 
computation. The magnetic fluxes across the cells are computed with a piecewise linear
interpolation method and the fluxes are combined with a Lax-Friedrichs Riemann
solver, with a time integration based on the total
variation diminishing second order Runge-Kutta scheme \citep[][]{wa09}. 
Thanks to the capabilities of \texttt{ENZO} of selectively refining interesting patches in the domain at higher resolution, we used adaptive mesh refinement (AMR) to
selectively increase the dynamical resolution in the formation region of galaxy
clusters and groups, which is necessary to properly resolve structure formation and
overcome the effect of magnetic field dissipation in converging flows at low
resolution \citep[][]{va14mhd}. 

\begin{figure}[t]
\centering
\includegraphics[width=\columnwidth]{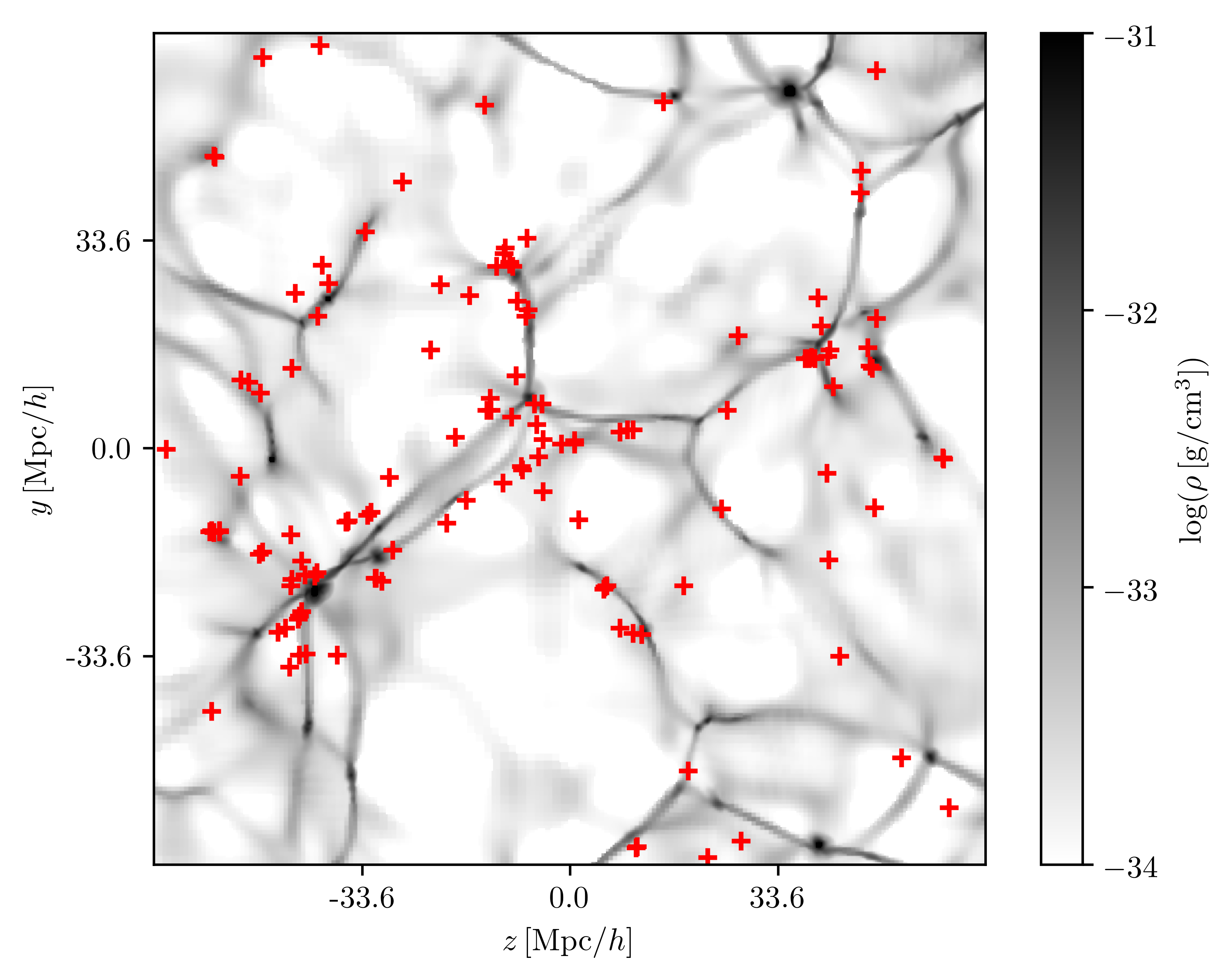}
\caption{\label{fig:2mpp}A slice through the gas density distribution of the innermost region of the box averaged over $6$ voxels in $x$ direction  at redshift $z=0$ as a result of the \texttt{ENZO} simulation. The plane is about $1\,\mathrm{Mpc}\,h^{-1}$ thick. The red crosses indicate the positions of galaxies found by the {\tmpp} survey in the same volume.}
\end{figure}

In detail, we apply AMR only in the innermost $(120\, \mathrm{Mpc}\,h^{-1})^3$ region of the
simulation, centred on the Milky Way location, and allowed for 5 levels of
refinement (by increasing the resolution of a factor 2 at each level, therefore up
to a $2^5=32$ refinement) whenever the local gas/dark matter density exceeded the mean density of
surrounding cells by a factor of 3; the procedure is recursively
repeated at each AMR level. This ensures that the magnetic field evolution in the
innermost clusters regions is typically followed with a spatial resolution of
$61-122\, \mathrm{kpc}\,h^{-1}$ (comoving) within the innermost AMR region of our volume. 
To confirm the consistency of our result, we show a slice through the gas density resulting from the simulation in Fig. \ref{fig:2mpp}. This plot indicates that we reproduce the large scale structure consistently with observations. 

\section{Reconstructing primordial magnetic fields}
\label{sec:results}

We will present the results of our work in two steps. First we focus on the statistical properties of the field at recombination. By applying the procedure described in the previous sections to an ensemble of data constrained initial conditions we can propagate observational uncertainties of the  matter distribution as traced by the {\tmpp} survey to the derived magnetic fields. In doing so we arrive at an ensemble of initial magnetic fields which constitutes a numerical description of the magnetic field posterior distribution at redshift $z = 1088$ conditional on {\tmpp} galaxy data. The goal here is to show how these uncertainties translate onto the calculated primordial magnetic field and to give scale dependent estimates on correlations and field strengths at this epoch.

The second part will show the results after the MHD run at redshift $z=0$. Here we will also turn our face on one particular realisation of the primordial magnetic field. We will show the large scale primordial magnetic field of some clusters of galaxies as well as the field in the close proximity to Earth. The resulting fields are available for download at \footnote{\url{https://wwwmpa.mpa-garching.mpg.de/~ensslin/research/data/data.html} or \doi{10.5281/zenodo.1190925}}.

\subsection{Recombination}

\subsubsection{Means and Variances}
\label{subsec:sig}

\begin{figure*}
\begin{minipage}[t]{0.49\linewidth}
        \centering
        \includegraphics[width=\linewidth]{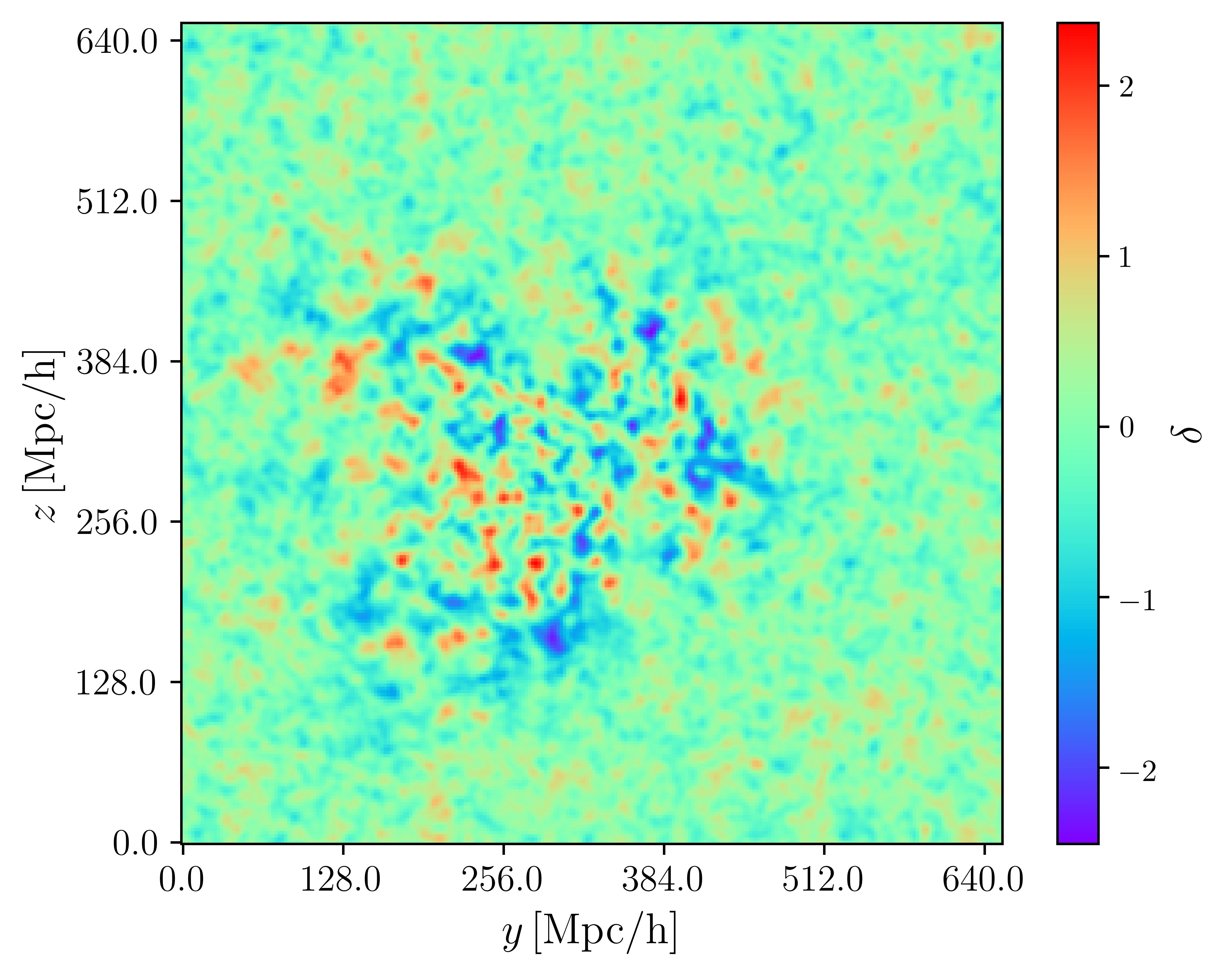}
    \end{minipage}%
    \hfill
\begin{minipage}[t]{0.49\linewidth}
        \centering
        \includegraphics[width=\linewidth]{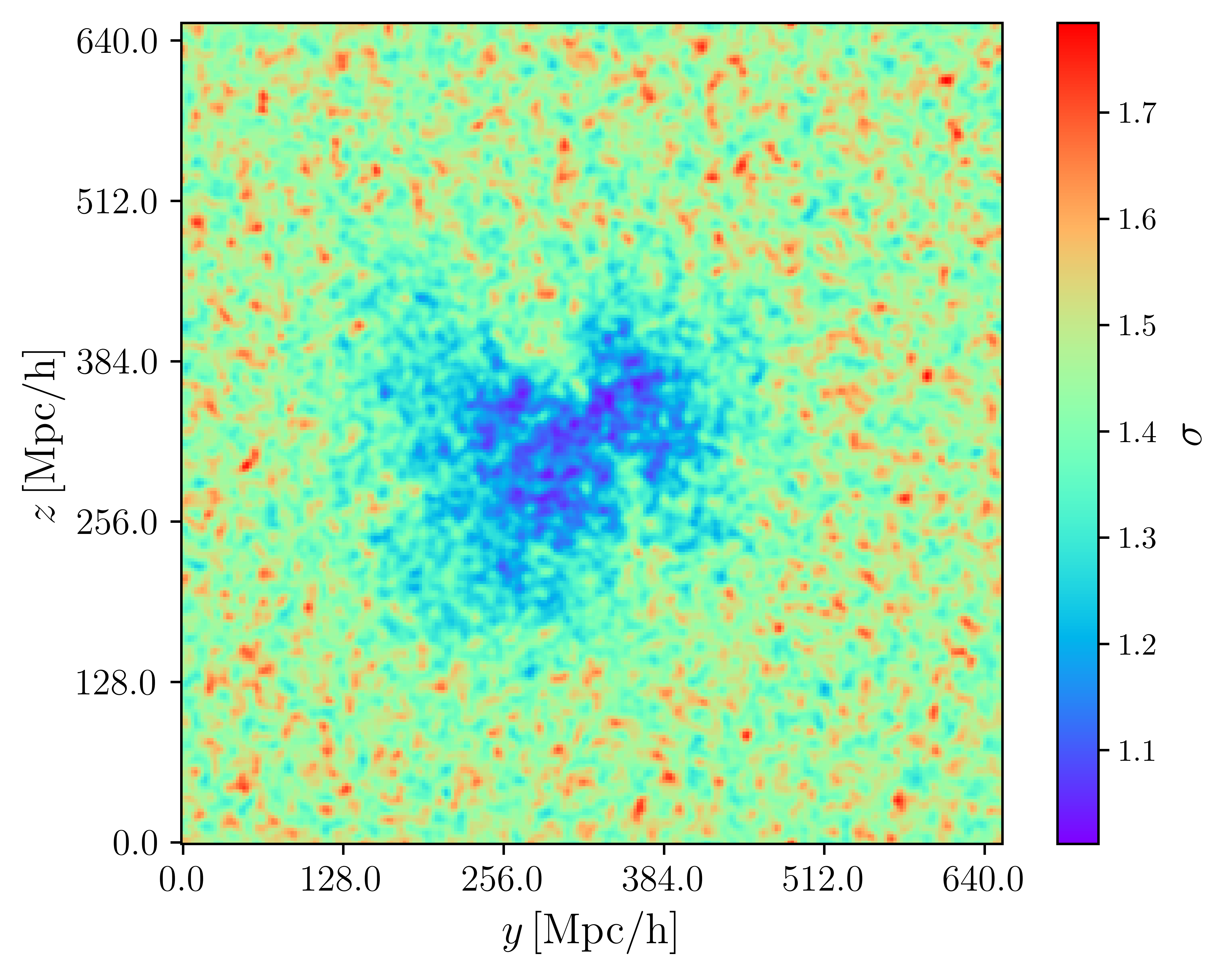}
    \end{minipage}
    \caption{\label{fig:borg} The posterior mean (left) and uncertainty standard deviation field of the dark matter overdensities (right) at redshift $z=1000$. This is the mean of the input data for our calculation averaged over 351 posterior samples of the matter field as generated by {\borg}. Our galaxy is centered in the middle. Areas close to the center a very pronounced in the mean, while areas further away are blurred out during the averaging. This reflects the Bayesian nature of the {\borg} algorithm, as the closer areas are very constrained by data, which leads to a relatively narrow posterior distribution in each pixel as reflected by the uncertainty variance. Therefore each sample looks similar there. The outer regions are barely constrained by data, leading to high uncertainties in the posterior.}
\end{figure*}

\begin{figure*}
\begin{minipage}[t]{0.49\linewidth}
        \centering
        \includegraphics[width=\linewidth]{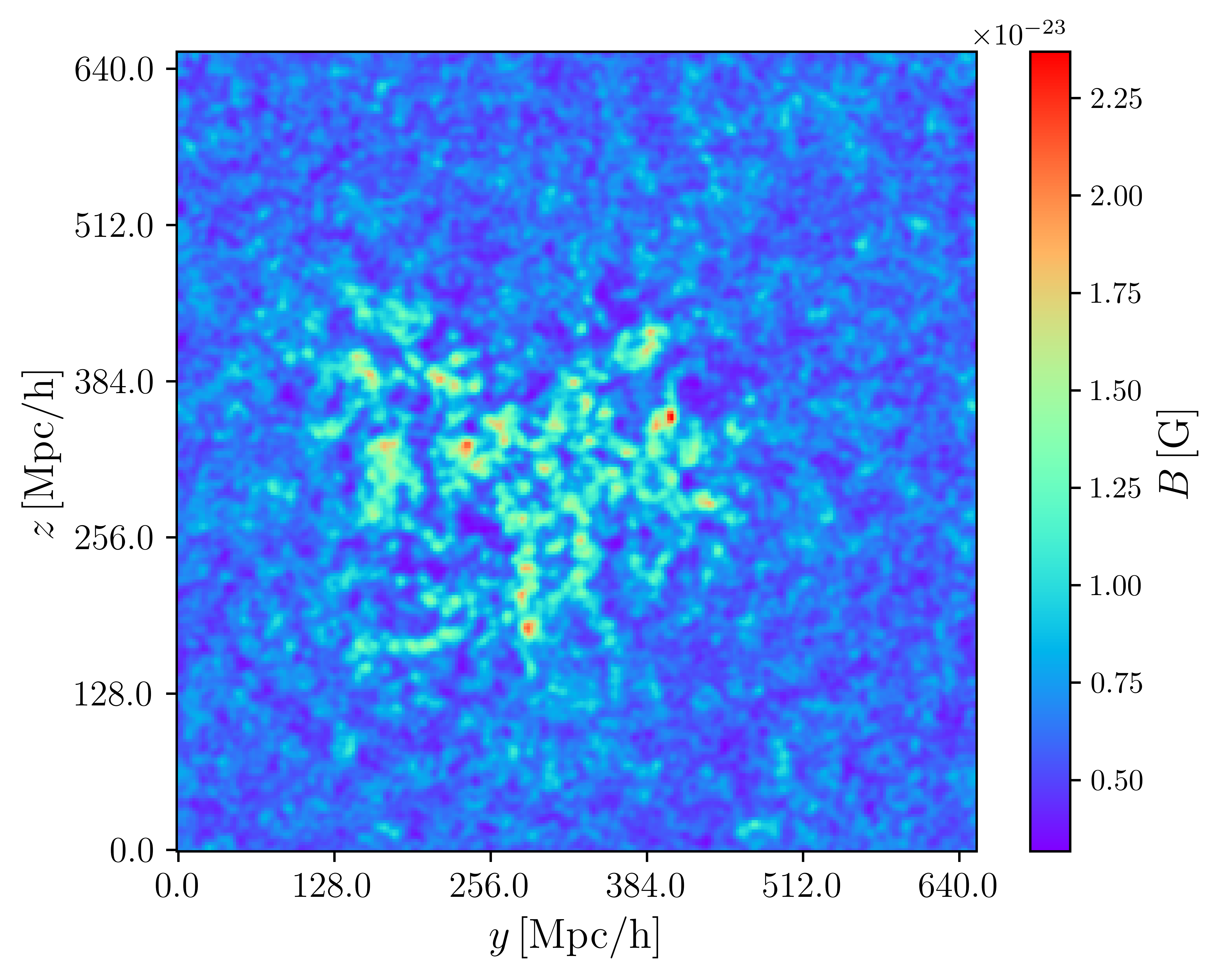}
    \end{minipage}%
    \hfill
\begin{minipage}[t]{0.49\linewidth}
        \centering
        \includegraphics[width=\linewidth]{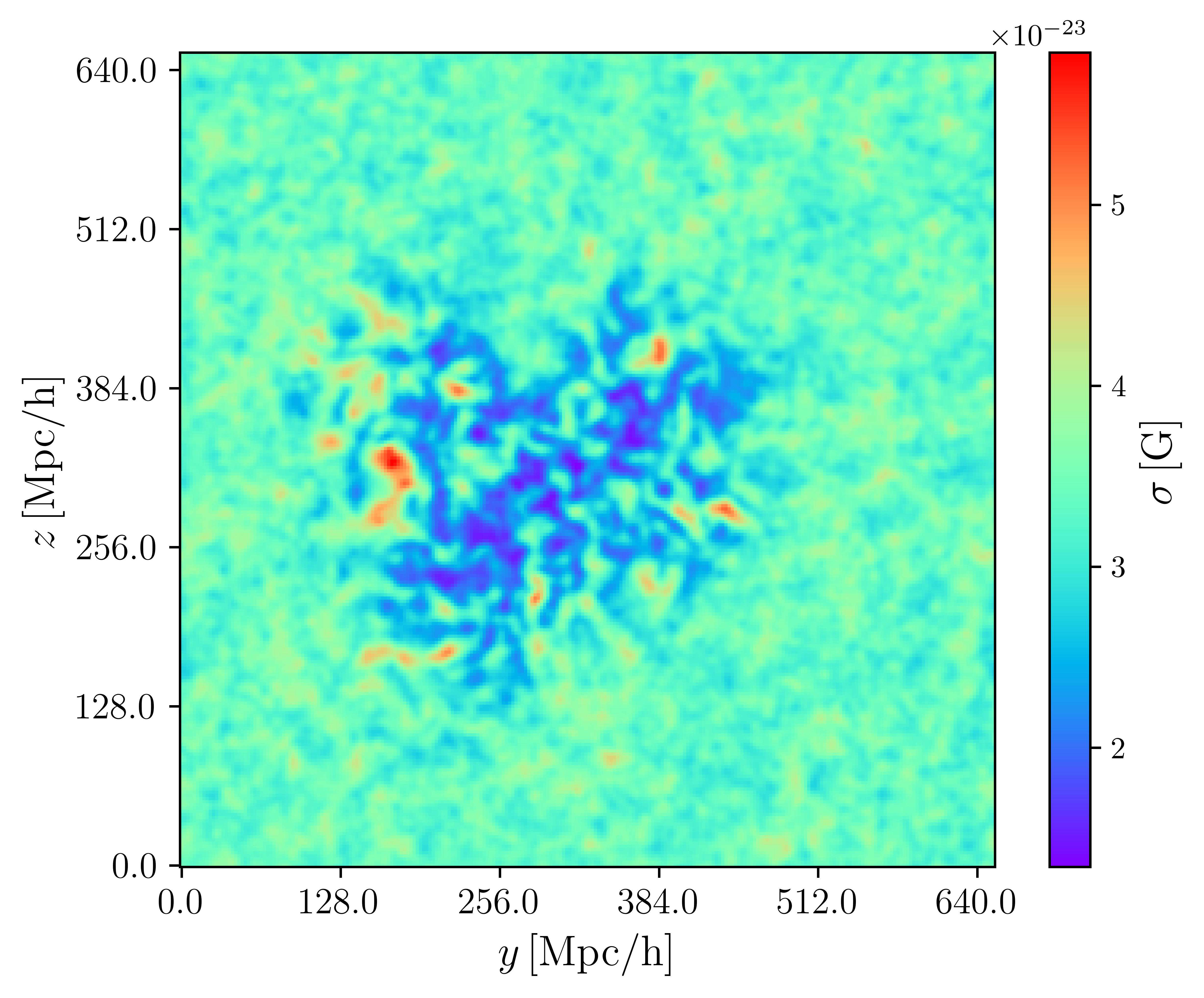}
    \end{minipage}
     \caption{\label{fig:B}The posterior mean (left) and uncertainty standard deviation (right) field of the absolute value of the Harrison magnetic field at redshift $z=1088$. Just as in the case of the initial data in Fig. \ref{fig:borg}, we note a very similar pattern in the mean and variance plots for regions closer and further away from Earth. The uncertainties of the density fields translate into uncertainties of the magnetic field.}
\end{figure*}

To illustrate the uncertainties we show slice plots of the input data and the resulting magnetic field strength at recombination in Figs. \ref{fig:borg} and \ref{fig:B}. All plots are slices through the $(677.7\, \mathrm{Mpc}\,h^{-1})^3$ cube. The first plot shows the field resulting form one particular sample of the {\borg} algorithm. The comoving root mean square (rms) field strength is around $10^{-23}\,\mathrm{G}$. The uncertainties are rather large compared to the mean. This is a consequence of he sparse data, which rather constrains the large scales than the small ones. Structures in the field appear to be rather small, typically with $\mathrm{Mpc}$-scale (see Section \ref{subsec:power}).  \par
Figs.  \ref{fig:borg} and \ref{fig:B} give an impression of the Bayesian properties of the {\borg} algorithm, which is translated onto our magnetic field realisations. They show the posterior mean and variance field of the magnetic field strength generated from 351 samples from the {\borg} posterior distributions. Areas which are highly constrained by data have well distinguishable structures in the mean, and have low uncertainty variance. The outer regions are less constrained, structures which are well visible in one particular sample are averaged out in the mean, and the variance is high.   

\subsubsection{Power spectra} 
\label{subsec:power}

Information on the correlation structure of a scalar field $s(x)$ can be gained from the corresponding scalar power spectrum defined as:

\begin{equation}
\label{eq:power}
\langle s(k)s^\ast(k^\prime)\rangle_{P(s)} = (2\pi)^3\, \delta(k-k^\prime)\,P_s(k)  
\end{equation}
where the asterisk denotes the complex conjugate.
In case of a magnetic field $B$, the statistically isotropic  and homogeneous correlation tensor is defined as

\begin{equation}
\label{eq:power_v}
\langle B_i(k) B_j^\ast(k^\prime)\rangle_{P(B)} = (2\pi)^3\, \delta(k-k^\prime)\,M_{ij}(k), 
\end{equation}
where the tensor $M_{ij}$ is defined as (see e.g. \citep{durcap}):

\begin{equation}
\label{eq:power_tensor}
M_{ij} = \frac{1}{2}\left(\delta_{ij} - \hat{k}_i\hat{k}_j\right)P_\mathrm{B}(k) + i \epsilon_{ijk}\hat{k}_kP_\mathrm{H}(k). 
\end{equation}

The helical part $P_\mathrm{H}(k)$ is assumed to be zero in this work. The magnetic field power spectra are therefore just the trace component of the magnetic field power spectrum tensor.

\begin{figure*}
\begin{minipage}[t]{0.49\linewidth}
\centering
\includegraphics[width=\linewidth]{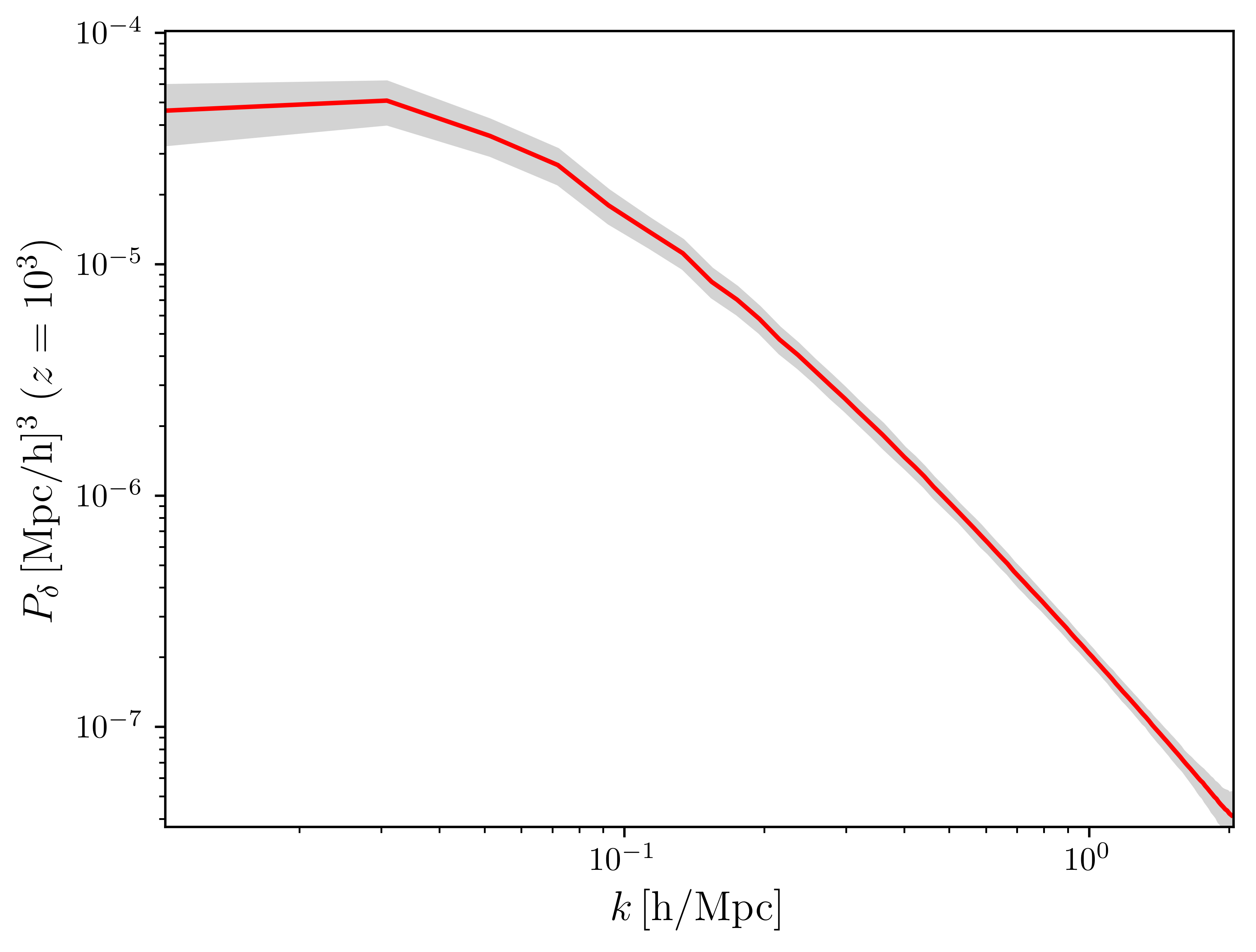}
\caption{\label{fig:matter_power} The matter power spectrum at $z = 10^3$. This is the spectrum of the input data. The red line is the mean averaged over the 351 samples. The grey area gives the uncertainty in the spectrum.}
\end{minipage}%
    \hfill
\begin{minipage}[t]{0.49\linewidth}
        \centering
\includegraphics[width=\linewidth]{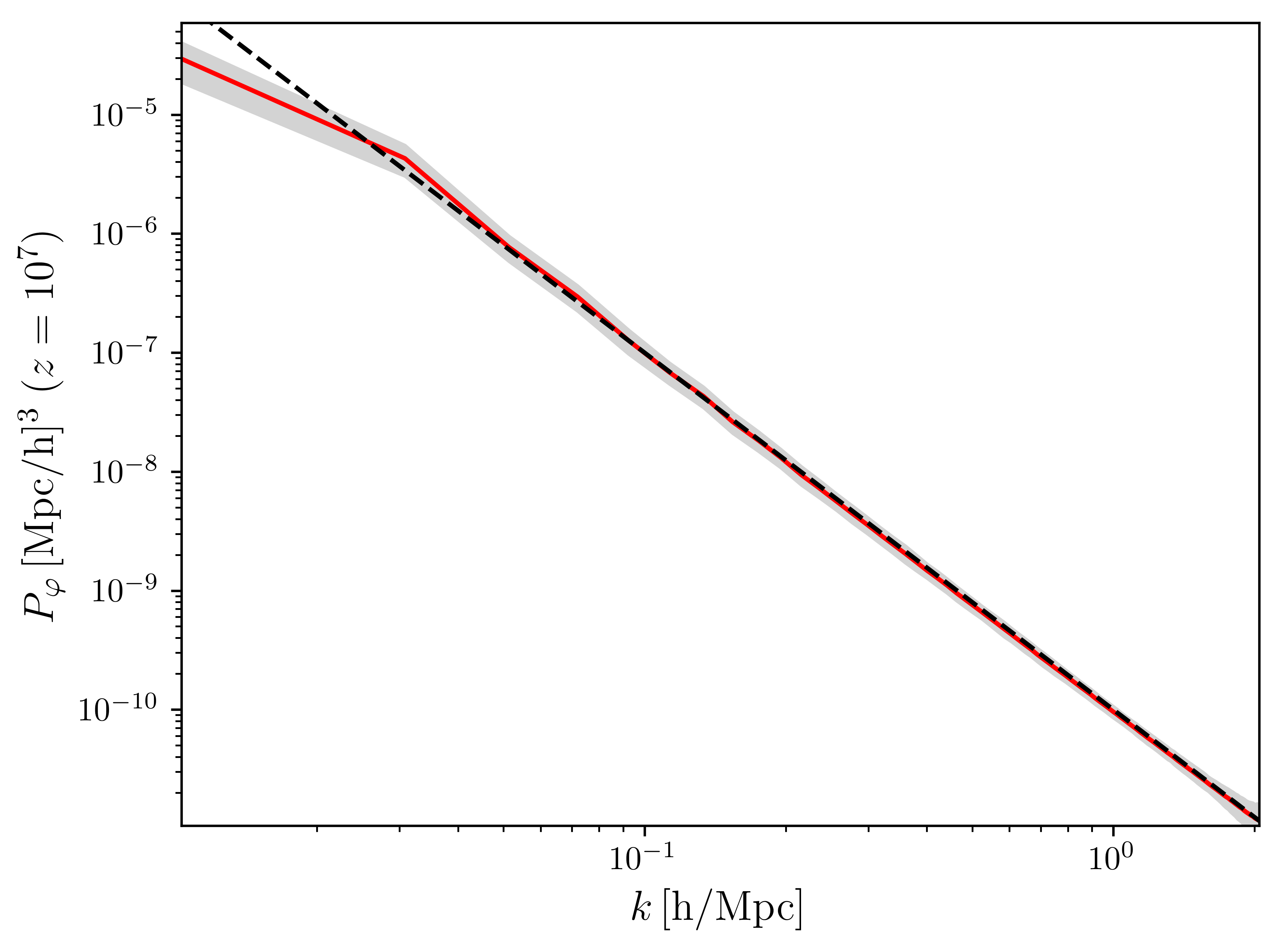}
\caption{\label{fig:potential_power_prim} The power spectrum of the primordial scalar perturbations at redshift $z=10^7$ as extracted from the cosmic structure reconstruction by \cite{Jasche14}. The dashed black line indicates the scale invariant spectrum normalized with the Planck amplitude parameter $\mathcal{A}_s$, see Tab. \ref{tab:imp_prim}.}
    \end{minipage}
\end{figure*}

\begin{figure*}
\begin{minipage}[t]{0.49\linewidth}
\centering
\includegraphics[width=\linewidth]{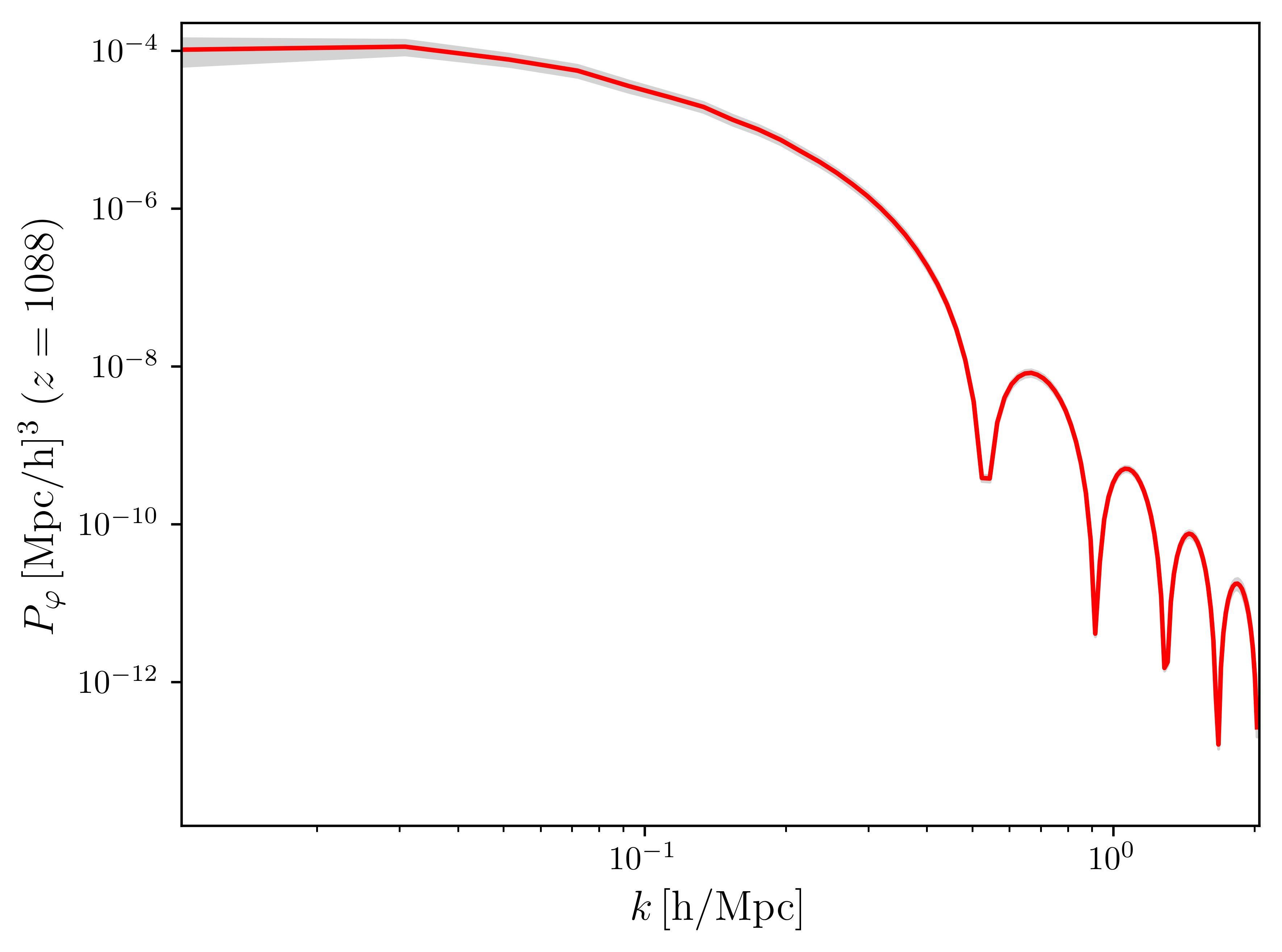}
\caption{\label{fig:potential_power_eq} The power spectrum of the scalar perturbations at redshift $z=3402$ at the end of the radiation dominated epoch. The oscillations in the spectrum are the Baryon-Accoustic Osclillations (BAO).}
\end{minipage}%
    \hfill
\begin{minipage}[t]{0.49\linewidth}
        \centering
\includegraphics[width=\linewidth]{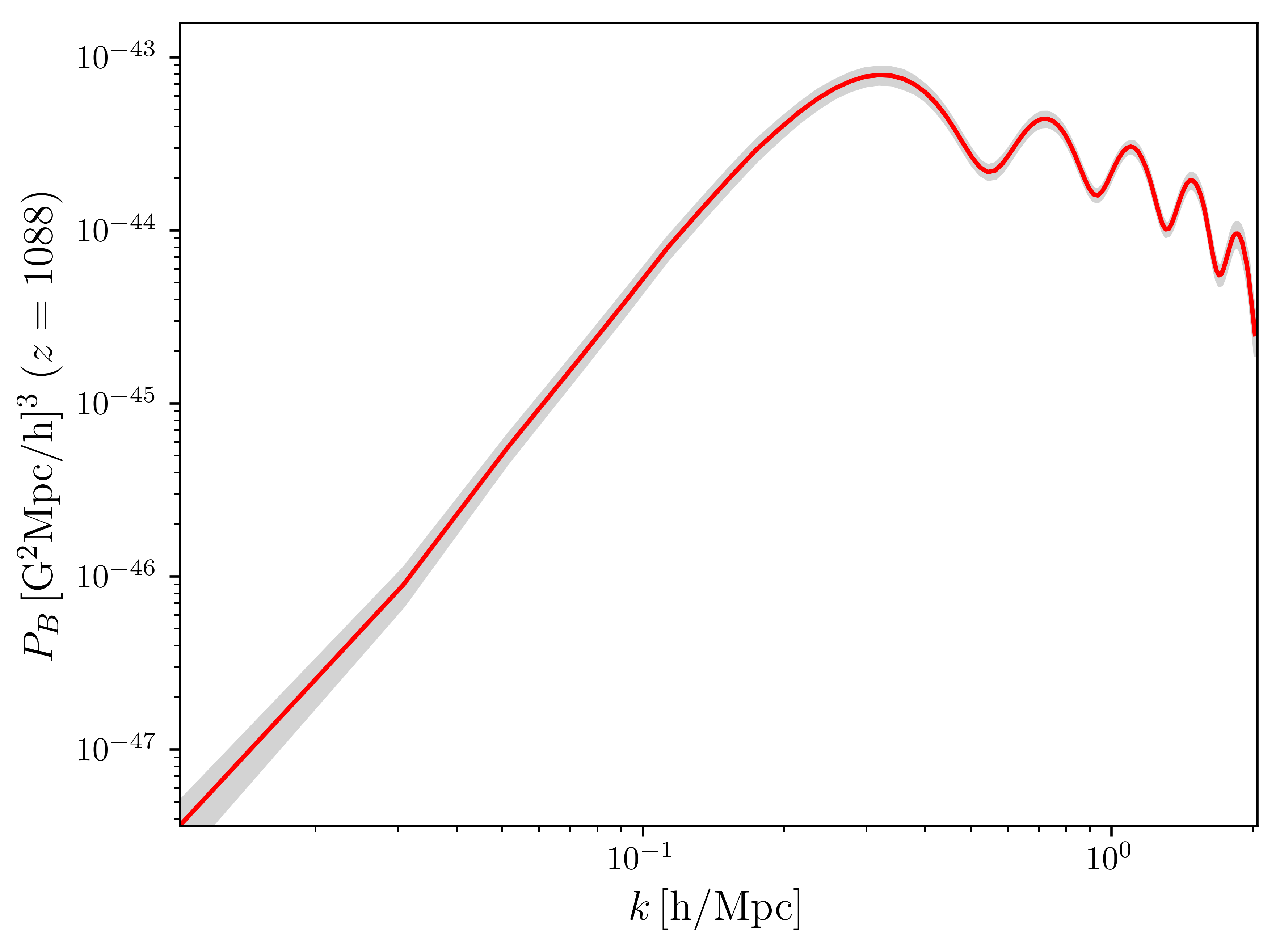}
\caption{\label{fig:B_power} The power spectrum of the magnetic field at redshift $z=1088$ just before recombination. The spectrum is defined of a vector field is defined in Eq. \eqref{eq:power_v} and Eq. \eqref{eq:power_tensor}. The spectrum peaks at approximately $\approx 3\cdot 10^{-1}\,\mathrm{Mpc}^{-1}\,h$.}
    \end{minipage}
\end{figure*}

As a consistency check we first show the power spectrum of the initial CDM field and the scalar perturbations $\varphi$ through some of the time steps of the algorithm in Figs. \ref{fig:matter_power}, \ref{fig:potential_power_prim} and \ref{fig:potential_power_eq}. These plots show the averaged spectra from 351 samples from {\borg} together with the corresponding uncertainties. We also show the magnetic field power power spectrum in Fig. \ref{fig:B_power}.

Despite some deviations on very large scales reflecting the uncertainties mentioned in the previous section, the spectra agree to a very good level with our expectations. These deviations can be noted in our intitial matter fields coming from the \texttt{BORG} algorithm (Fig. \ref{fig:matter_power}) and further on in all the other averaged power spectra.
We note a clear $k^{-3}$ dependence in the primordial potential and matter power spectrum corresponding to an approximately unity spectral index as expected for uncorrelated and scale invariant structures \citep{harrison_spectrum, zeldovich}. We can compare the spectrum with the Planck results as a consistency check, see Tab. \ref{tab:imp_prim} and the dashed line in Fig. \ref{fig:potential_power_prim}.

\begin{figure}
\begin{tabular}{|c|c|c|}
\hline
& Planck(2015) & This work \\
\hline
$\ln\left(10^{10} \mathcal{A}_s\right)$ & $3.064 \pm 0.023$ &  $\approx 3.1$\\
\hline
$n_s$ & $0.9667 \pm 0.0040$ & $\approx 1$ \\ 
\hline
\end{tabular}
\caption{\label{tab:imp_prim} Comparison of inflation parameters provided by the Planck collaboration \citep{Planck15_inflation} and as inferred from the samples used in this work.}
\end{figure}

The potential power spectrum at matter-radiation equality drops shortly above $k = 0.1\,\mathrm{Mpc}^{-1}\,h$, indicating the size of the horizon at that time. At small scales, the spectrum shows oscillations in Fourier space, which stem from the functional form of the potential evolution equation in Eq. \eqref{eq:phi_eq}. Physically speaking, these are the Baryon-accoustic oscillations (BAO's) induced by horizon crossing during the radiation epoch. The uncertainties again agree with the initial dark matter spectrum.  
The resulting power spectrum of the magnetic field is plotted in Fig. \ref{fig:B_power}. It  rises for small $k$-values with approximately $k^{3.5}$ as expected from our discussion in Sec. \ref{sec:theory} and peaks at $k_{\mathrm{peak}}\approx 2\cdot 10^{-1}\,\mathrm{Mpc}^{-1}\,h$. 
The plot shows little 'bumps' on small scales, which are remnants of the oscillating potential in the radiation epoch. 
At this point it shall also be noted that in the time frame of our calculation any turbulence due to primordial velocity perturbations is not relevant. In \citep{banjeree} the authors show that given these perturbations the very Early Universe has Reynolds numbers in the range of $10^3$. This then gives the perfect framework for a small-scale dynamo to amplify the magnetic seed fields originating from the Harrison effect to fields with strengths of approximately $10^{-15}\, \mathrm{G}$, but with typical correlation lengths of the order of parsecs. Given the Mpc resolution of our grid, this is not relevant for this work.

\subsubsection{Scale dependent mean field}
\label{subsec:scale}

To give a more intuitive picture of the expected magnetic field strengths, we convolve the magnetic field power spectrum with a Gaussian kernel in position space to get an estimate for $B$ given a scale of reference $\lambda$.

\begin{equation}
\label{eq:convolution}
B_{\lambda}^2 =\frac{1}{(2\pi)^3} \int P_B(k)\,e^{-\frac{k^2\lambda^2}{2}} d^3k
\end{equation}

\begin{figure}[t]
\centering
\includegraphics[width=\columnwidth]{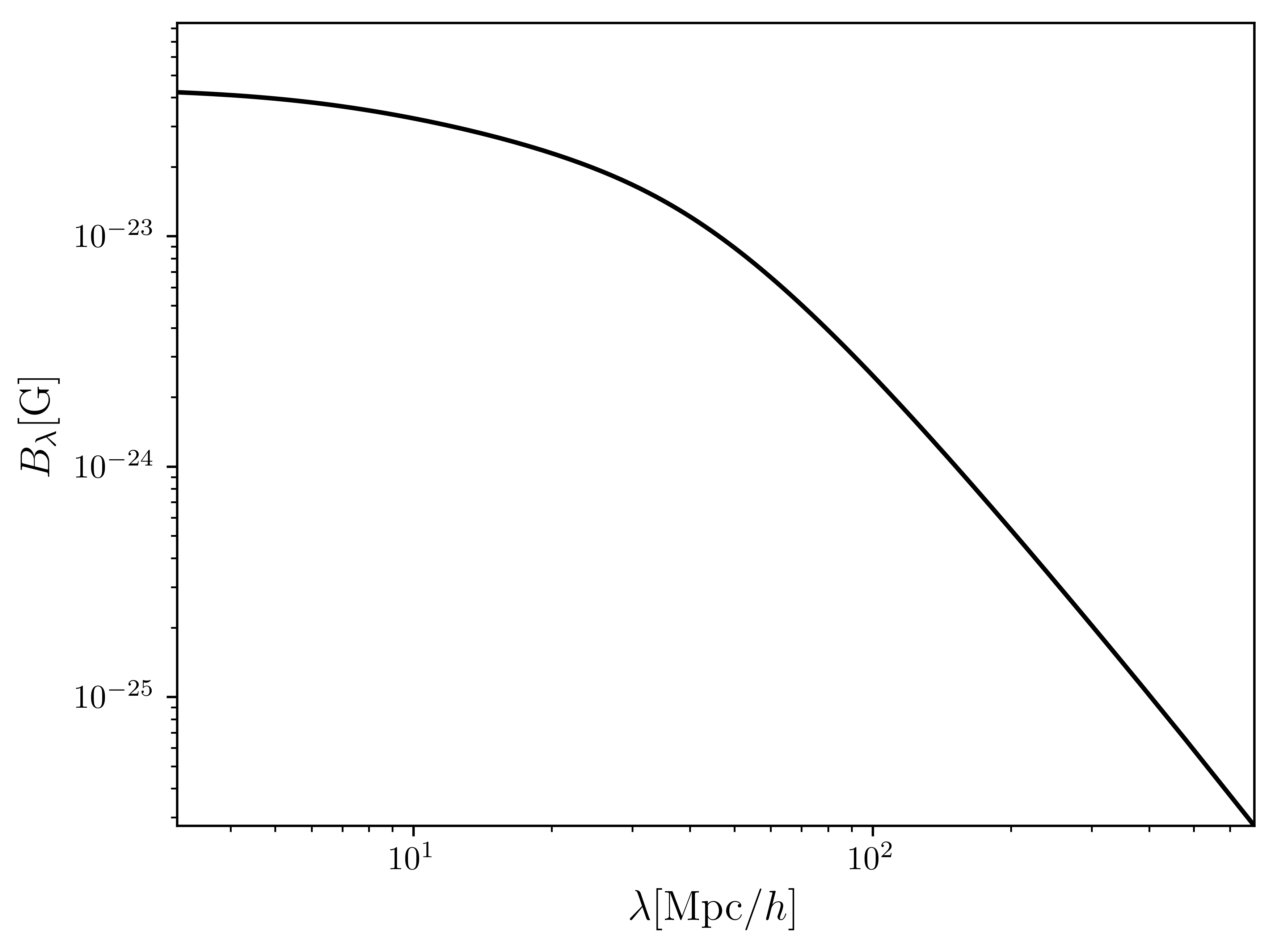}
\caption{\label{fig:Bmeanlambda}Scale averaged magnetic field at recombination. This is the result of Eq. \eqref{eq:convolution}.}
\end{figure}

The result of this operation is shown in Fig. \ref{fig:Bmeanlambda}. For scales reaching from $2.65\, \mathrm{Mpc}\,h^{-1}$  to  $\approx 10\, \mathrm{Mpc}\,h^{-1}$ the magnetic field strength weakly declines and has a typical strength of approximately $10^{-23}\, \mathrm{G}$. 
For scales larger than $10\, \mathrm{Mpc}\,h^{-1}$, $B_\lambda$ roughly scales as
\begin{equation}
B_\lambda \sim \lambda^{-2.5}.
\end{equation} 
The field strength reaches from $10^{-23}\, \mathrm{G}$ at the smallest scales just over $1\,\mathrm{Mpc}\,h^{-1}$ to less than $10^{-27}\, \mathrm{G}$ at scales over a $100\,\mathrm{Mpc}\,h^{-1}$. This information is of course closely related to the magnetic field power spectrum.

\subsection{Today}

\subsubsection{Field strength and correlation structure}
\label{subsec:corr}

\begin{figure*}
\begin{minipage}{0.49\linewidth}
\centering
\includegraphics[width=\linewidth]{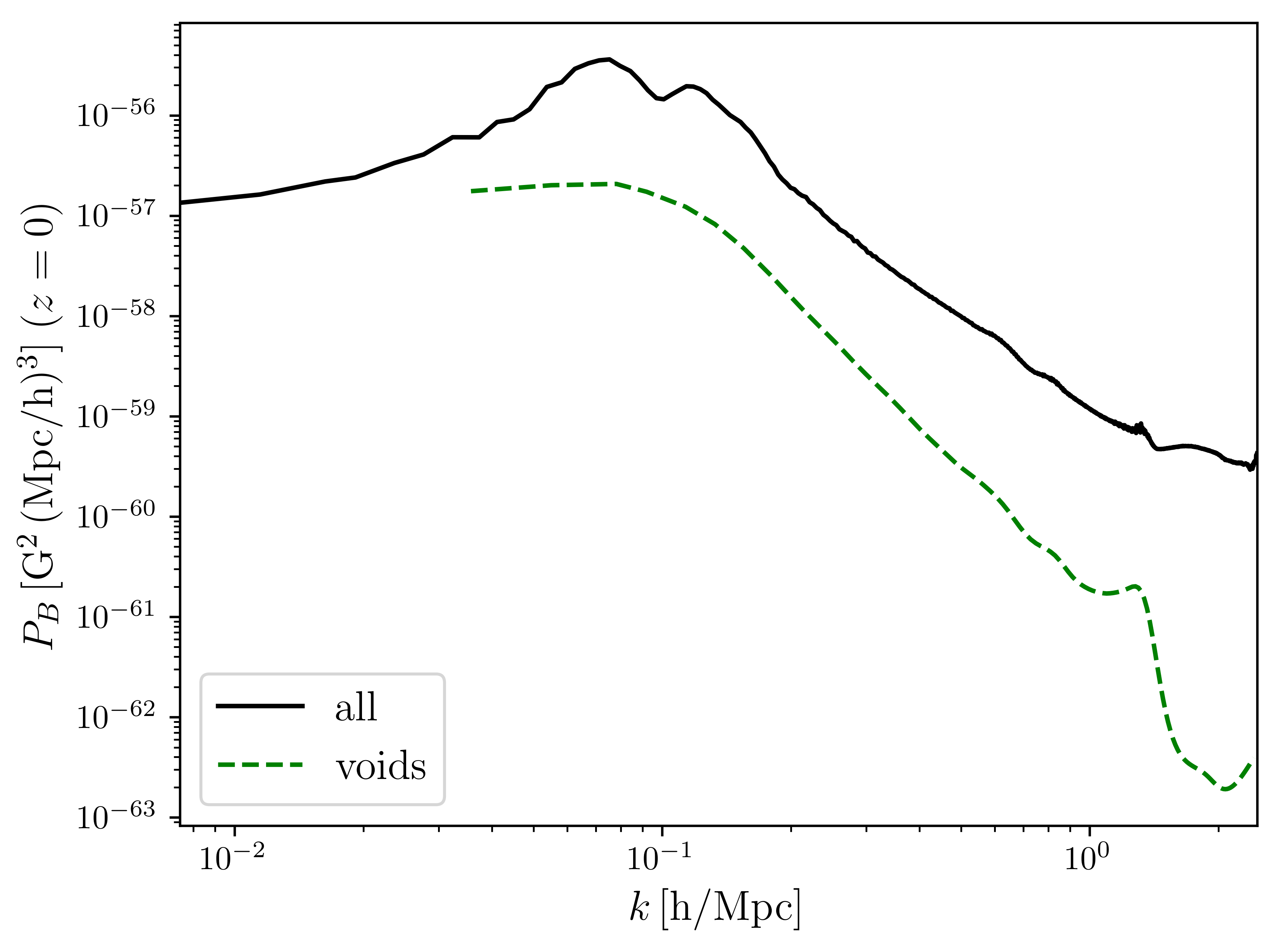}
\caption{\label{fig:magnetic_power} The magnetic power spectrum at $z = 0$, defined according to Eq. \eqref{eq:power_v}. The black line indicates the spectrum for the complete magnetic field in the box. The green dashed line indicates the void power spectrum, computed only from a part of the box. Here we considered voxels with gas density $\rho < 3\cdot \overline{\rho}$ as void voxels. The void power spectrum was inferred using the critical filter technique \citep{jakob}, which assumes that the unmasked regions are typical for the whole volume.}
\end{minipage}
    \hfill
\begin{minipage}{0.49\linewidth}
        \centering
\includegraphics[width=\linewidth]{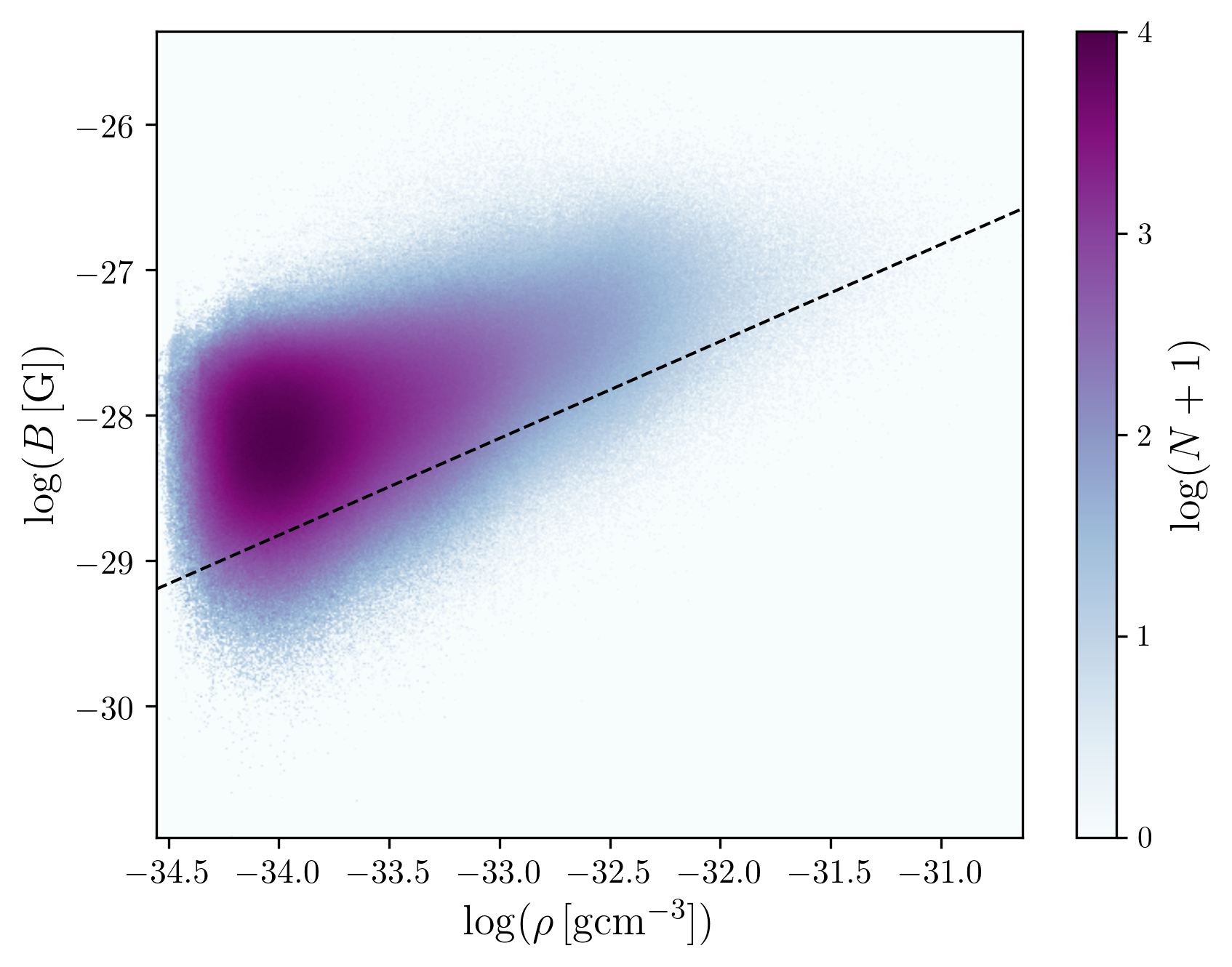}
\caption{\label{fig:scatter_0} Joint histogram of the magnetic field and matter density at redshift $z=0$ with $512^2$ bins. The dashed line indicates the $B\propto \rho^{2/3}$ relation resulting from the flux freezing of the magnetic field lines. This relation is also observed in simulations starting with unconstrained magnetic field conditions, see e.g.\citep{va17}.}
\end{minipage}
\end{figure*}

In Figs. \ref{fig:magnetic_power} we depict the power spectrum of the magnetic field today for one sample of the \texttt{BORG} posterior. We show the complete spectrum as well as the void power spectrum inferred via negleting dens voxels with gas density $\rho > 3\cdot \overline{\rho}$ and the critical filter technique \citep{jakob}, which assumes that the unmasked regions are typical for the whole volume. The BAO signature and most small structures have been destroyed during structure formation, leading to a mostly red spectrum. For the complete spectrum and the voids, most power still lies on scales of about $k_{\mathrm{peak}}\approx 10^{-1}\,\mathrm{Mpc}^{-1}\,h$. The morphology of the complete power spectrum is rather similar to the void power spectrum, which is expected, as they compromise the largest volume share of the Universe and calculating a power spectrum is effectively a volume averaging procedure. The decrease at large scales again reflects the solenoidality of magnetic fields ($\nabla \cdot B= 0$) for uncorrelated signals \citep{durcap}, as the large scale structure has a characteristic size and therefore larger scales are not strongly causally connected via gravity. \par
In Fig. \ref{fig:scatter_0} we show the joint probability function of matter density and magnetic field strength. Most of the probability mass lies on rather small densities, with varying magnetic field strengths. Large densities tend to be associated with large magnetic field strength. The lower bound of this plot follows a $B\propto \rho^\frac{2}{3}$ proportionality, which was already found in previous simulations, e.g: \citep{va17}. All in all this leads to the picture that the magnetic field in the low density areas which are relatively little affected by structure formation mostly retain their correlation structure and morphology. After recombination, the field is frozen into the plasma. Therefore the field strength scales with $a(t)^{-2}$, explaining the field strengths somewhere around $10^{-29}\,\mathrm{G}$. Within dense structures the field is at least amplified up to $10^{-26.5}\,\mathrm{G}$. We underline this view with specific examples in the next chapter.

\subsubsection{Field realisations}
\label{subsec:field}

\begin{figure*}
\begin{minipage}{\linewidth}
\centering
\includegraphics[width=\linewidth]{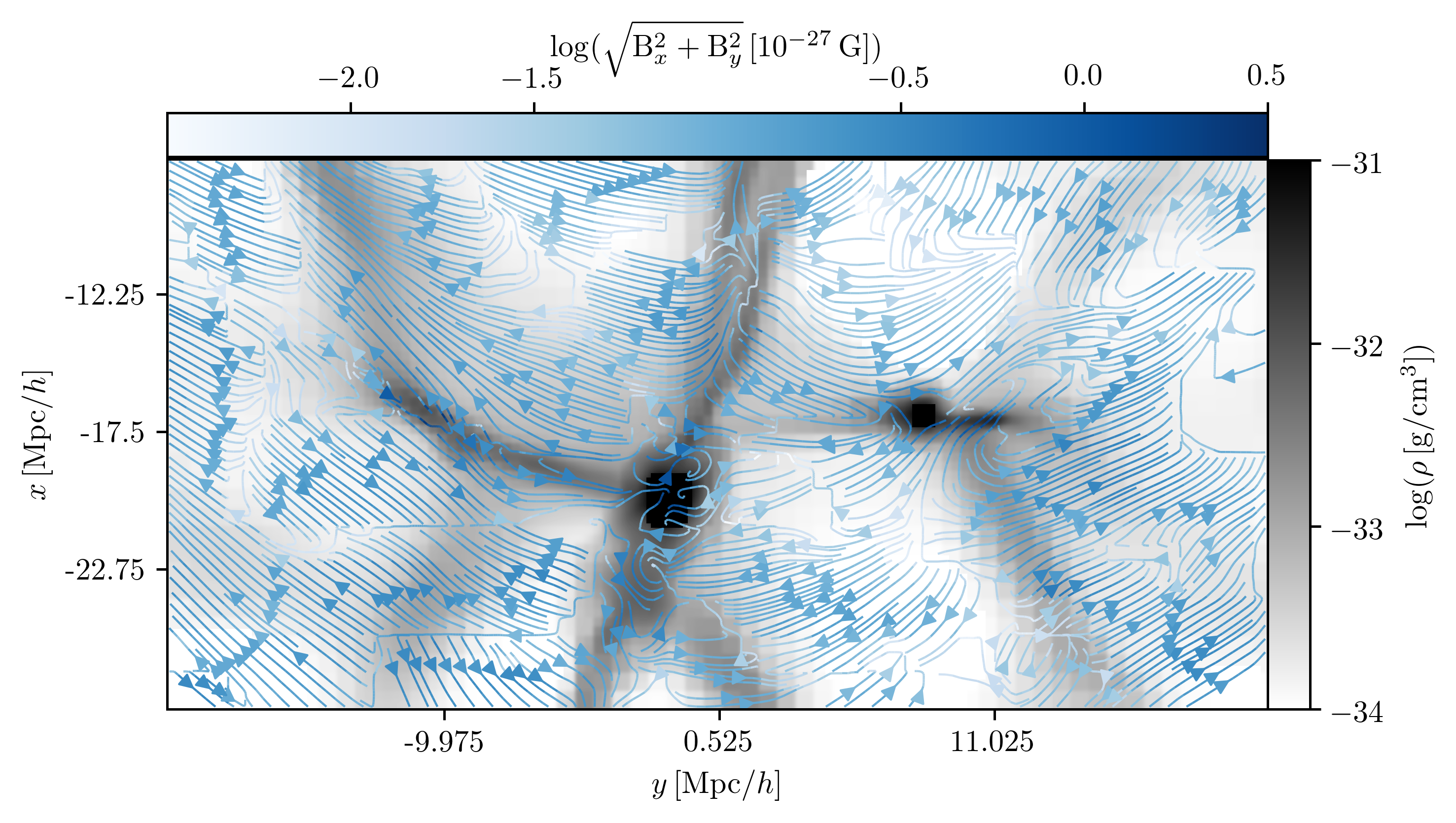}
\end{minipage}%

\begin{minipage}{\linewidth}
\centering
\includegraphics[width=\linewidth]{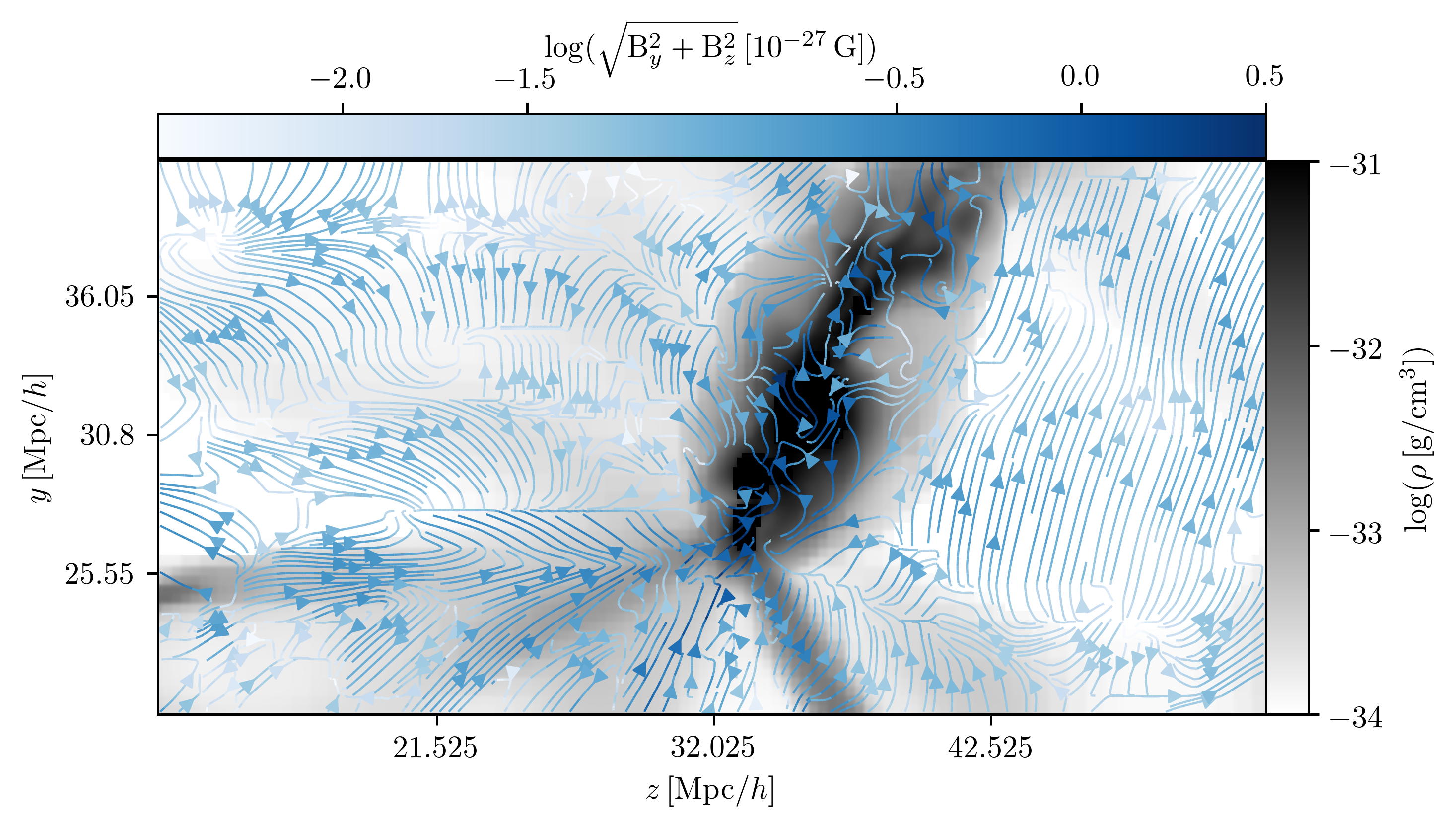}
\end{minipage}
\caption{\label{fig:clusters} The magnetic field and gas matter density in a slice trough the Virgo (above) and the Perseus-Pisces (below) cluster. The plots shows the gas matter density overplotted with the $y-z$ components of the magnetic field vectors. All colorbars have a logarithmic scaling. The coordinates are defined via the equatorial plane with reference to the galactic centre. The choice of the slice is purely for artistic reasons.}
\end{figure*}

\begin{figure*}
\begin{minipage}{\linewidth}
\centering
\includegraphics[width=\linewidth]{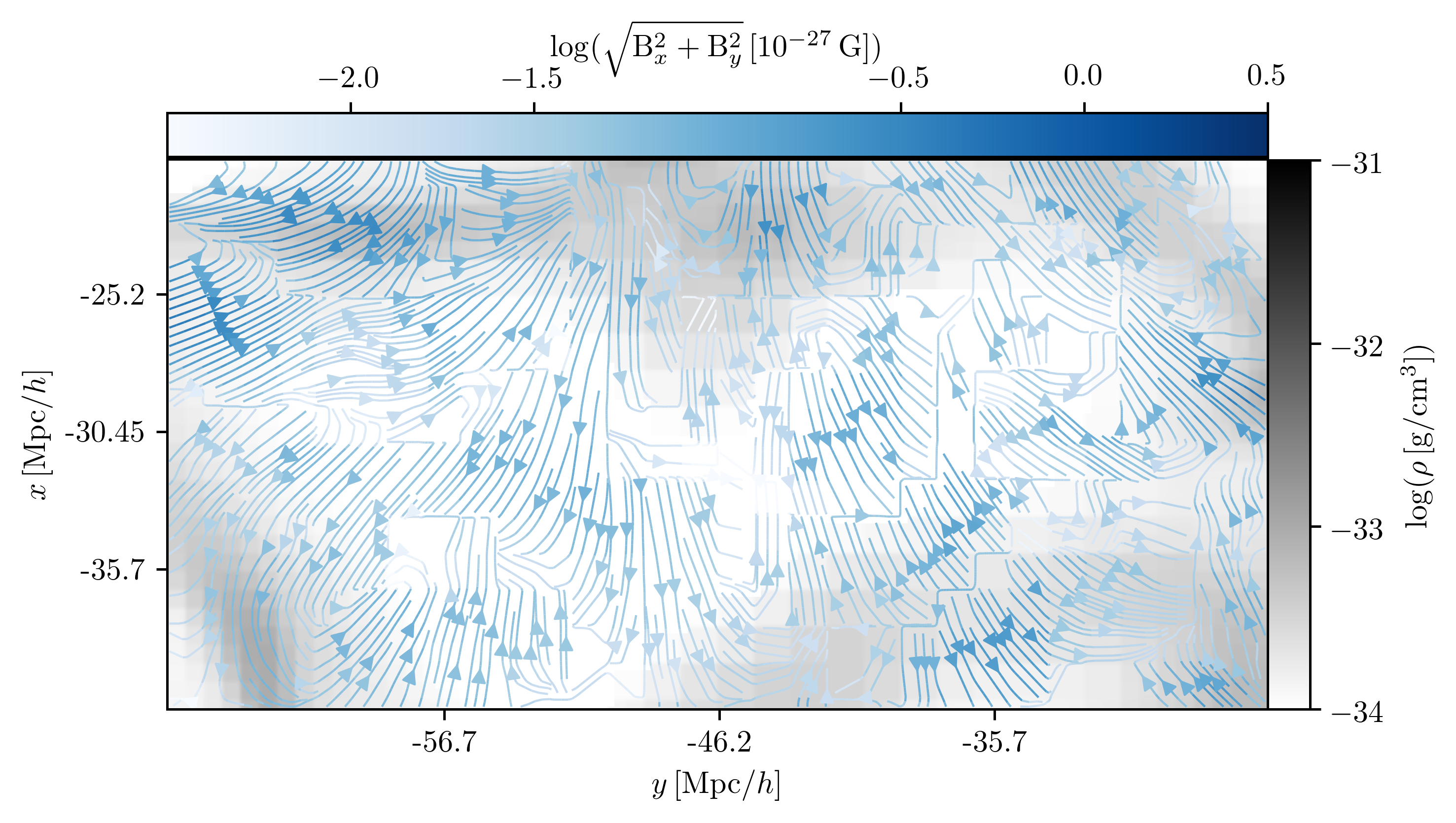}
\end{minipage}%

\begin{minipage}{\linewidth}
\centering
\includegraphics[width=\linewidth]{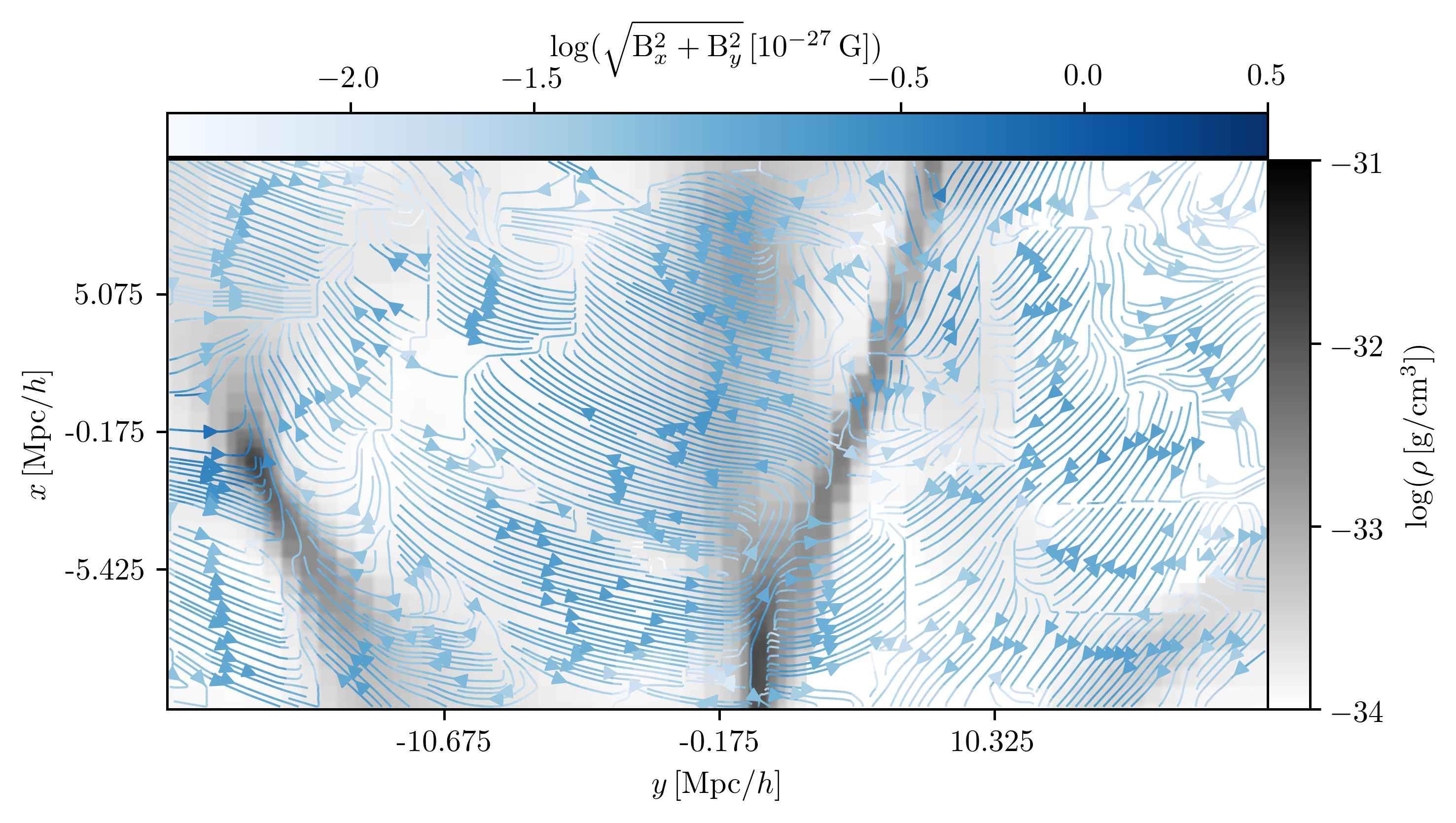}
\end{minipage}
\caption{\label{fig:voidhome} The magnetic field and gas matter density in an underdense region (above) and around the galactic center in the $x-y$ plane. The plots shows the gas matter density overplotted with the $x-y$ components of the magnetic field vectors. All colorbars have a logarithmic scaling. The coordinates are defined via the equatorial plane with reference to the galactic centre.}
\end{figure*}

In Figs.\ref{fig:clusters} and \ref{fig:voidhome} we show structures in the magnetic field and the density field which belong to different morphological features of the Local Universe. We find that the magnetic field strength strongly correlates with the gas density in all of these structures, consistent with a frozen-in behaviour of magnetic fields. In very dense clusters such as Virgo and Perseus-Pisces in Fig. \ref{fig:clusters} the magnetic field morphology seems to be driven by the infall of matter on the cluster. Of course the simulation is too coarse to correctly cover the structure formation and magnetic field behaviour on small scales within these structures, for this reason any small scale structures in these plots are highly uncertain. Also the maximum magnetic field strength maybe higher, as we cannot resolve any potential dynamo mechanism during structure formation. In underdense regions as depicted in the upper image in Fig. \ref{fig:voidhome}, we observe a morphology similar to the initial conditions, with a charcteristic scale of a few $\mathrm{Mpc/h}$. Apart from the aforementioned $a(t)^{-2}$ dependence of the field strength, the morphology seems to relatively unaffected, which is consistent with our view of a 'frozen' magnetic field. In the lower image of Fig. \ref{fig:voidhome} we show the magnetic field in a slice around our galaxy. The field here is slightly amplified up to field strengths of $10^{-28}\, \mathrm{G}$, as a slight overdense structure seems to have formed in the region, which may correspond to the Local Group. 
The Local Group has a typical scale of about $2$ Mpc, which is slightly below the smallest data constraint scale in our calculation, making the association difficult.

\subsubsection{Full sky maps}
\begin{figure*}
\centering
\includegraphics[scale=0.65]{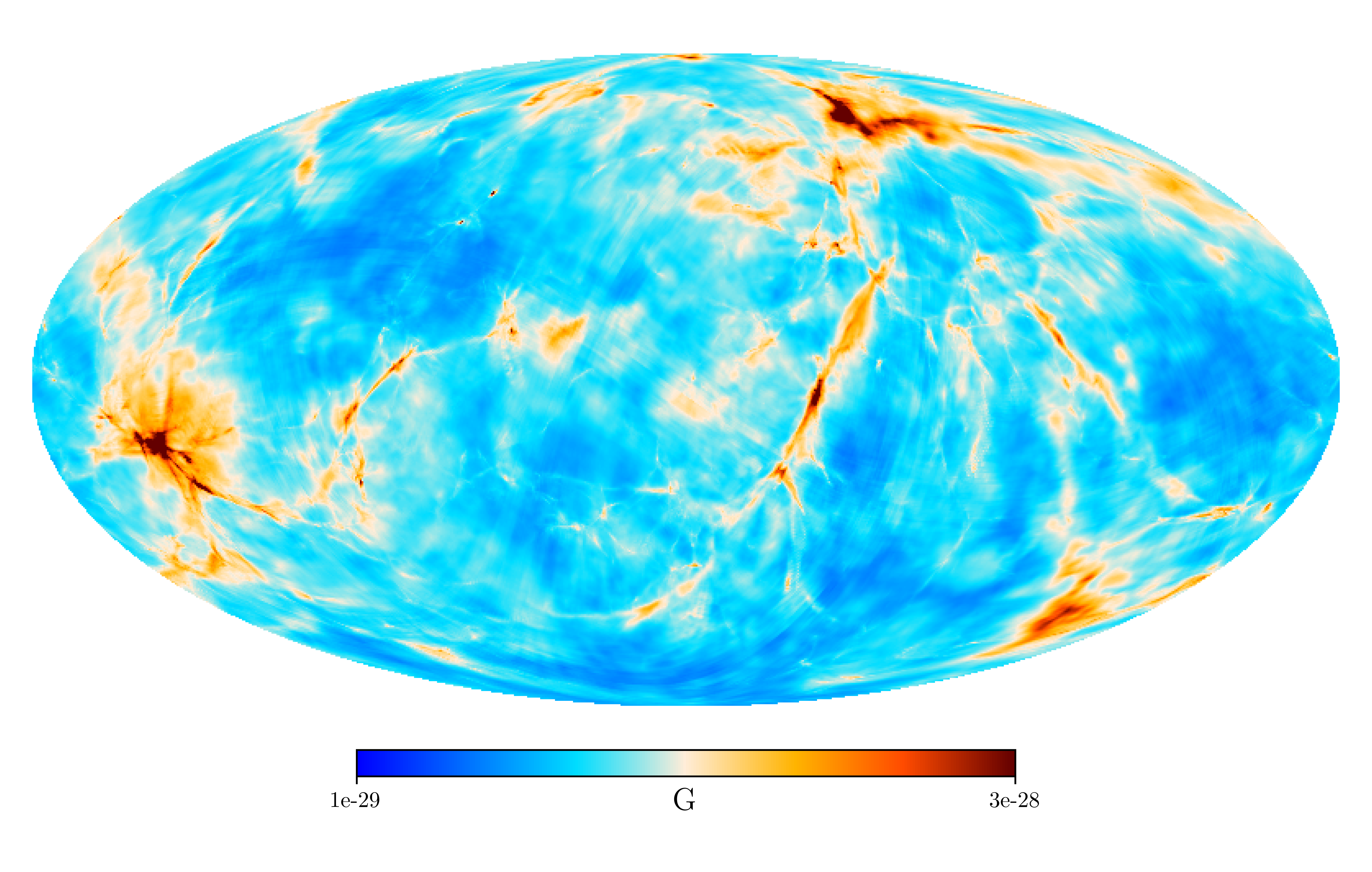}
\caption{\label{fig:B_LOS} The magnetic field strength averaged over line of sights in units of Gauss for sources within a distance of 60 Mpc/$h$ from Earth. The plot is in galactic coordinates. The two dominant clusters in this image are Persues Pisces in the middle left of the image and Virgo close to the North pole. Close ups of both structures are provided in Fig. \ref{fig:clusters}.} 
\end{figure*}

\begin{figure*}
\centering
\includegraphics[scale=1.1]{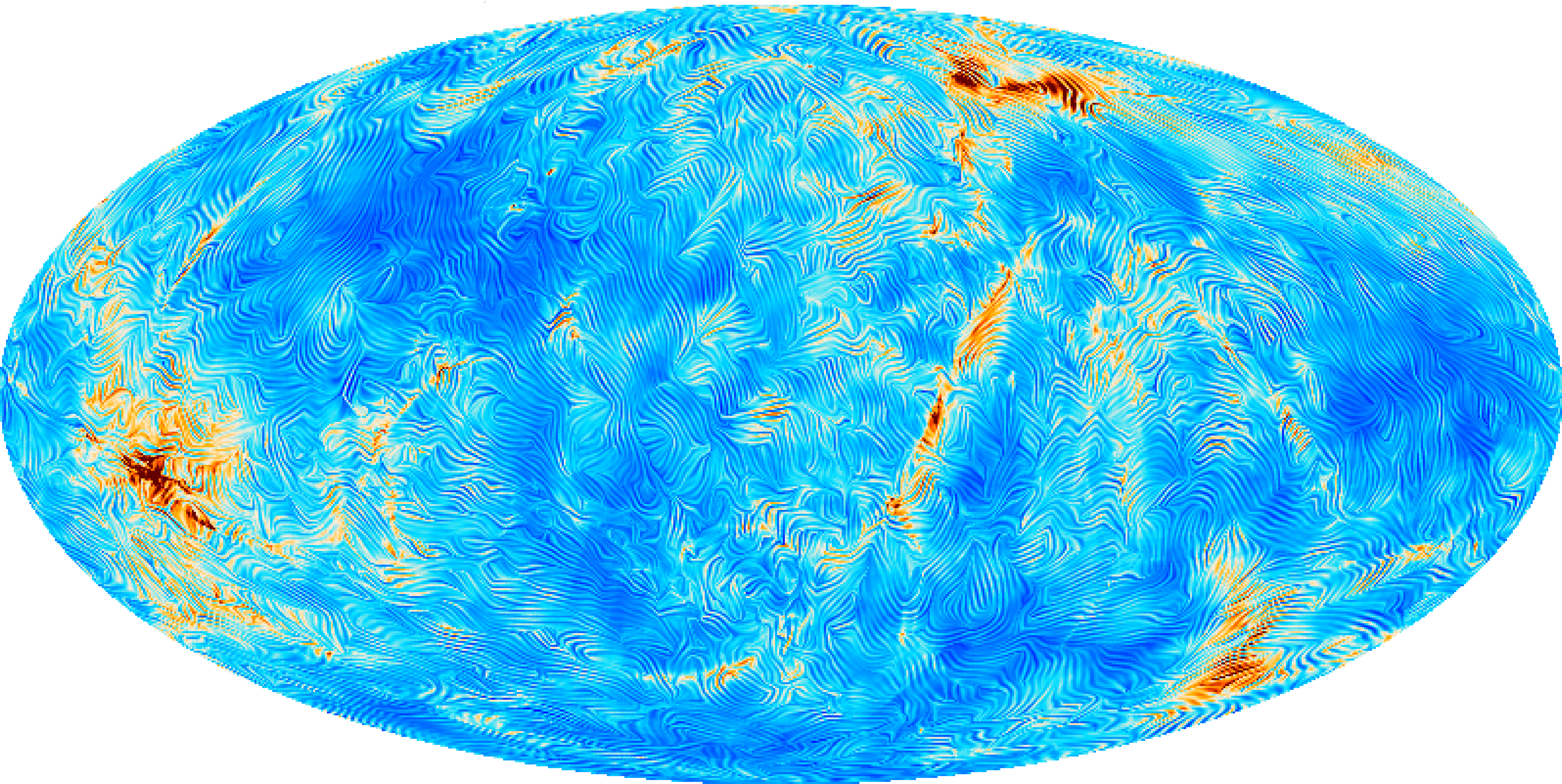}
\caption{\label{fig:Bnewpol} A polarization-like plot visualizing the magnetic field morphology perpendicular to the LOS. This plot was generated using the 'Alice' module of the HEALPix software\footnote{\url{http://healpix.sourceforge.net}} \cite{healpix} and the linear integral convolution algorithm \citep{cabral}. The plot is in galactic coordinate.} 
\end{figure*}

\label{subsec:faraday}
\begin{figure*}
\centering
\includegraphics[scale=0.65]{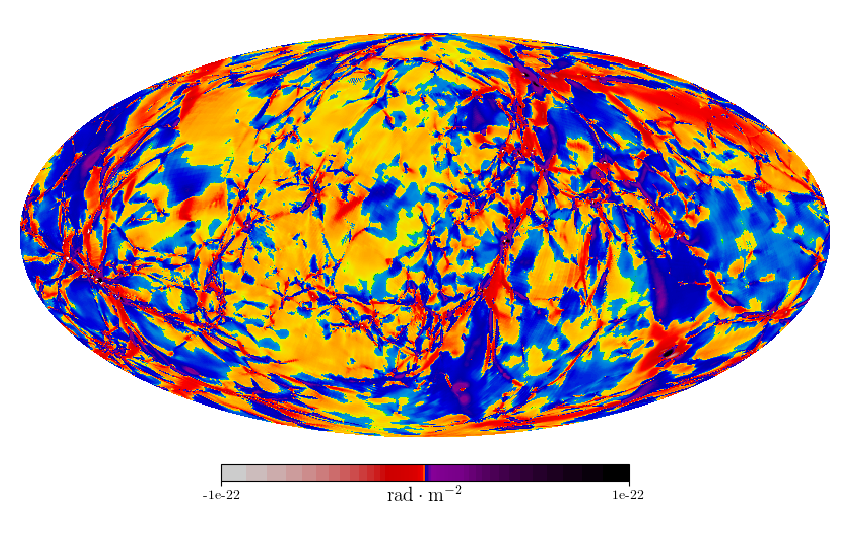}
\caption{\label{fig:faradayplot}  The primordial magnetic field Faraday rotation measure for polarized sources located within a distance of 60 Mpc$h^{-1}$ from earth in units of radians per square metre. The plot is in galactic coordinates. The colormap is logarithmic on both the negative and the positve regime with a linear scaling between $-10^{-29}$ and $10^{-29}$ rad$ \cdot \mathrm{m}^{-2}$, connecting both parts of the scale. We used the rescaled gas mass density as an estimate for the electron number density.}

\centering
\includegraphics[scale=0.65]{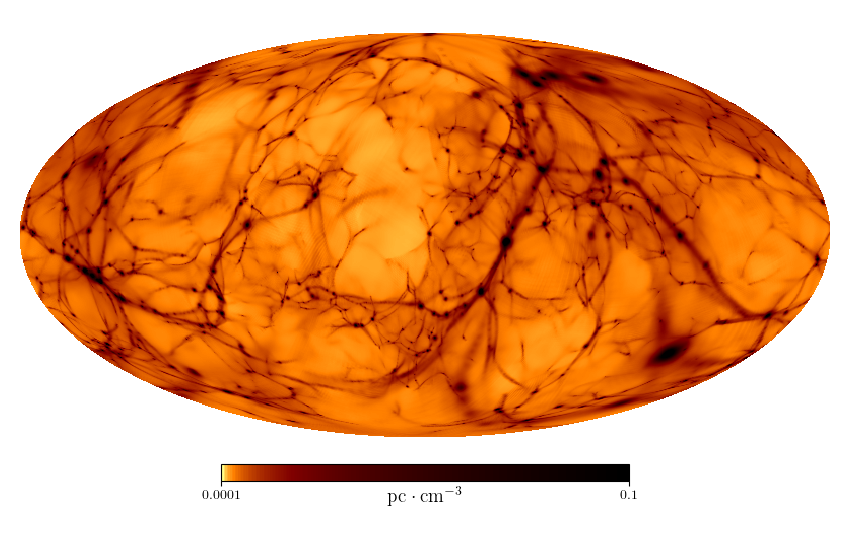}
\caption{\label{fig:dmplot} The electron dispersion measure in units of parsecs per cubic centimetre for sources within a distance of 60 Mpc/$h$ from Earth. The plot is in galactic coordinates. We used the rescaled gas mass density as an estimate for the electron number density. The two dominant clusters in this image are Persues Pisces in the middle left of the image and Virgo close to the North pole. Close ups of both structures are provided in Fig. \ref{fig:clusters}.} 
\end{figure*}

We can use the results of the \texttt{ENZO} simulation to estimate the expected Faraday rotation of linear polarized light under the influence of a magnetic field. 
Faraday rotation measure (RM) is calculated via 

\begin{equation}
\label{eq:faraday}
\mathrm{RM} = \frac{e^3}{2\pi m_e^2c^4} \int_0^{R_{\mathrm{max}}} n_{\mathrm{th}}\, B\, dr
\end{equation}
in cgs units (see e.g. \citep{oppermann}). It is essentially a line of sight (LOS) integration up to a distance $R_{\mathrm{max}}$ over the magnetic field $B$ weighted with the electron number density $n_{\mathrm{th}}$.
This can be computed using the publicly available Hammurabi Software \citep{hammurabi}, which performs the necessary LOS integration over a sphere around the Earth\footnote{We used a reimplementation of Hammurabi [Wang et al., in prep.]; https://bitbucket.org/hammurabicode/hamx}. The result can be seen in Fig. \ref{fig:faradayplot}. The same software is also able to calculate the dispersion measure of electrons (DM) and the LOS averaged absolute magnetic field strength, 
\begin{equation}
\label{eq:dm}
\mathrm{DM} = \int_0^{R_{\mathrm{max}}} n_{\mathrm{th}}\, dr,
\end{equation} 
shown in Figs. \ref{fig:dmplot} and \ref{fig:B_LOS}. These maps nicely trace dense structures in the sphere over which we integrated. We also used the output of Hammurabi for the LOS perpendicular components of the magnetic to generate a polarization like plot in Fig. \ref{fig:Bnewpol}, which traces the magnetic field morphology in the sphere.  Comparing the plots we see that areas of large RM correspond to large electron densities, as we expect given the linear $n_{\mathrm{th}}$ dependence of RM. We also note again that the magnetic field strength and morphology correlates with the density.\par
Of course the expected signal is beyond any chance of measurability, and in the realistic case we expect that the memory of any such tiny seed field within clusters is entirely lost due to the dynamo amplification process \citep{bermin}, which is expected to be much more efficient on scales smaller than what is resolved at our resolution here. 
The void signal, however, although a few order of magnitudes smaller, may be relatively undisturbed by such processes, at least away from other possible sources of magnetisation, like dwarf galaxies \citep{beck}.

\section{Summary and Discussion}
\label{sec:outlook}

We calculated the large scale primordial magnetic field originating from the Harrison effect \citep{harrison} and second order vorticity generation in the radiation dominated era. This is the first time a data constraint reconstruction of the remnant of a primordial magnetic field was achieved.\par
For that, we used our knowledge about the large scale structure in the Universe coming from the {\tmpp} galaxy survey and the {\borg} algorithm to infer the corresponding density distributions deep in the radiation epoch. Using an existing formalism for the magnetic field generation from these initial conditions, we then found at recombination a field coherent on comoving scales in the $10\, \mathrm{Mpc}\,h^{-1}$ regime, with a maximum field strength of about  $ 10^{-23}\, \mathrm{G}$ at these scales. By means of a MHD simulation we evolved the magnetic field through structure formation and came up with field strengths higher than  $\approx 10^{-27}\, \mathrm{G}$ and $\approx 10^{-29}\, \mathrm{G}$ in clusters of galaxies and voids, respectively. We specifically showed the structure of the field around well known structures in the Local Universe, such as the Virgo and Perseus Pisces cluster.\par
The above results, including the statistical properties of the magnetic fields, the morphology of the field on above Mpc scales and the expected
observables shown in Figs. \ref{fig:faradayplot} and \ref{fig:dmplot} rely only on the assumption of a $\Lambda\mathrm{CDM}$ cosmology and conventional plasma physics. We introduced further simplifications such as the tight coupling approximation and the simplified modelling around the radiation matter equality due to computational constraints. In \citep{fenu, Saga}, the authors calculated the correct evolution equations without these simplifications, leading to slightly different spectra, but comparable magnetic field strengths.
Large scale magnetic fields can also be produced by more speculative mechanisms for primordial magnetogenesis, by transferring magnetic energy of small scaled magnetic fields to larger scales via an inverse cascade and by magnetogenesis driven by radiation pressure during reionisation \citep{sub_io}. For this reason, we view our results as a lower limit on the magnetic field strength in the Local Universe.
This is especially true for clusters, as for once small scales are not strongly constraint by our data and moreover we where not able to resolve the relevant scales for magnetic field amplification via turbulence, as predicted by e.g. Subramanian et al. \citep{sub_cluster}.
We did arrive at magnetic field strengths which could act as a seed field for the galactic dynamo \citep{davis}, however given the fact that we cannot adequately resolve sub-Mpc scales and galactic magnetic fields maybe explained without a primordial seed, we refrain from giving an estimate to which extent the Harrison magnetic field could have influenced galactic magnetic fields.
A possible explanation for the non-observation of TeV-photons from blazars are void magnetic fields of strength $10^{-15}$ G \citep{neronov, neronov2}, among other explanations\citep{pfrommer}. If these fields exist, our prediction is not sufficient to explain them.\par
Considering the rather conservative assumptions made in our calculations, we can provide a credible lower bound on the strength of the large scale magnetic field today and an impression of its expected morphology. The logical next step building up on this work would be a refinement of the calculation via the implementation of more sophisticated formalisms for the generation of primordial magnetic fields, especially including a more accurate baryon photon interaction treatment.

\section*{Acknowledgments}
We thank four anonymous referees for insightful comments. This research was supported by the DFG Forschungsgruppe 1254 ``Magnetisation of Interstellar and Intergalactic Media:  The Prospects of Low-Frequency Radio Observations" and the
DFG cluster of excellence ``Origin and Structure of the
Universe" (www.universe-cluster.de). This work made
in the ILP LABEX (under reference ANR-10-LABX-63)
was supported by French state funds managed by the
ANR within the Investissements d’Avenir programme under reference ANR-11-IDEX-0004-02.
GL is supported in part
by the French ANR BIG4, grant ANR-16-CE23-0002. 
The cosmological simulations described in this work were performed using the \texttt{ENZO}
code (http://enzo-project.org), which is the product of a collaborative effort of
scientists at many universities and national laboratories. We gratefully acknowledge
the \texttt{ENZO} development group for providing extremely helpful and well-maintained
online documentation and tutorials.
The computation presented in this work was produced on at J{\"u}lich Supercomputing
Centre (JSC), under projects no. 10755 and 11764. F.V. acknowledges financial
support from the European Union's Horizon 2020 research and innovation programme
under the Marie-Sklodowska-Curie grant agreement no.664931,  and from the ERC
Starting Grant ``MAGCOW", no.714196. 
DP acknowledges financial support by the ASI/INAF Agreement I/072/09/0 for the
Planck LFI Activity of Phase E2 and by ASI Grant 2016-24-H.0.
Some of the results in this paper have been derived using the HEALPix \citep{healpix} package.

\clearpage
\bibliographystyle{unsrtnat}
\bibliography{lib}

\end{document}